# *OTFS: A New Generation of Modulation Addressing the Challenges of 5G*

## *Authors*

Ronny Hadani, Anton Monk

Cohere Technologies

## *Abstract*

In this paper, we introduce a new 2D modulation scheme referred to as OTFS (Orthogonal Time Frequency & Space) that multiplexes information QAM symbols over new class of carrier waveforms that correspond to localized pulses in a signal representation called the delay-Doppler representation. OTFS constitutes a far reaching generalization of conventional time and frequency modulations such as TDM and FDM and, from a broader perspective, it establishes a conceptual link between Radar and communication. The OTFS waveforms couple with the wireless channel in a way that directly captures the underlying physics, yielding a high-resolution delay-Doppler Radar image of the constituent reflectors. As a result, the time-frequency selective channel is converted into an invariant, separable and orthogonal interaction, where all received QAM symbols experience the same localized impairment and all the delay-Doppler diversity branches are coherently combined. The high resolution delay-Doppler separation of the reflectors enables OTFS to approach channel capacity with optimal performance-complexity tradeoff through linear scaling of spectral efficiency with the MIMO order and robustness to Doppler and multipath channel conditions. OTFS is an enabler for realizing the full promise of MU-MIMO gains even in challenging 5G deployment settings where adaptation is unrealistic.

# 1. OTFS – A NEXT GENERATION MODULATION

History teaches us that every transition to a new generation of wireless network involves a disruption in the underlying air interface: beginning with the transition from 2G networks based on single carrier GSM to 3G networks based on code division multiplexing (CDMA), then followed by the transition to contemporary 4G networks based on orthogonal frequency division multiplexing (OFDM). The decision to introduce a new air interface is made when the demands of a new generation of use cases cannot be met by legacy technology – in terms of performance, capabilities, or cost. As an example, the demands for higher capacity data services drove the transition from legacy interference-limited CDMA network (that have limited flexibility for adaptation and inferior achievable throughput) to a network based on an orthogonal narrowband OFDM that is optimally fit for opportunistic scheduling and achieves higher spectral efficiency.

Emerging 5G networks are required to support diverse usage scenarios, as described for example in [1]. A fundamental requirement is multi-user MIMO, which holds the promise of massive increases in mobile broadband spectral efficiency using large numbers of antenna elements at the base-station in combination with advanced precoding techniques. This promise comes at the cost of very complex architectures that cannot practically achieve capacity using traditional OFDM techniques and suffers performance degradation in the presence of time and frequency selectivity ( [2] and [3]). Other important use cases include operation under non-trivial dynamic channel conditions (for example vehicle-to-vehicle and high-speed rail) where adaptation becomes unrealistic, rendering OFDM narrowband waveforms strictly suboptimal. As a result, one is once again faced with the dilemma of finding a better suited air interface where the new guiding philosophy is:

**When adaptation is not a possibility one should look for ways to eliminate the need to adapt.**

The challenge is to do that without sacrificing performance. To meet this challenge one should fuse together two contradictory principles – (1) the principle of spreading (as used in CDMA) to obtain resilience to narrowband interference and to exploit channel diversity gain for increased reliability under unpredictable channel conditions and (2) the principle of orthogonality (as used in OFDM) to simplify the channel coupling for achieving higher spectral densities with a superior performance-complexity tradeoff.

OTFS is a modulation scheme that carries information QAM symbols over a new class of waveforms which are spread over both time and frequency while remaining roughly orthogonal to each other under general delay-Doppler channel impairments. The key characteristic of the OTFS waveforms is related to their optimal manner of interaction with the wireless reflectors. This interaction induces a simple and symmetric coupling between the channel response and the information carrying QAM symbols, thus facilitating the design of transmitter and receiver structures with optimal performance-complexity tradeoff. In summary, OTFS combines the reliability and robustness of spread spectrum with the high spectral efficiency and low complexity of narrowband transmission.

This paper consists of two parts. The first part is devoted to explaining the mathematical principles behind OTFS and the second part is devoted to demonstrating OTFS performance gains over multicarrier modulations like OFDM, focusing on the following core 5G use cases.

1. Enhanced Mobile Broadband (eMBB). In this use case we show that OTFS enables scaling of spectral efficiency with increased MIMO order under any channel condition with optimal performance-complexity tradeoff. We describe the principles of delay-Doppler equalization and precoding for MU-MIMO and its intrinsic advantage over the conventional time-frequency counterpart.
2. Internet of Things (IoT). In this use case we describe a specific OTFS transmission mode for small packets that maximizes the link budget (energy per bit) and minimizes the number of retransmissions under power and latency constraints, hence prolonging battery life and achieving extended coverage. The OTFS transmit signal enables low PAPR and maximum available duration (to maximize link budget) while extracting full time-frequency diversity (to minimize number of retransmissions).
3. Communication under high mobility conditions such as vehicle-to-vehicle communication (V2V) or high speed train (HST). In this use case we show that OTFS maximizes throughput, reliability, and performance consistency. We further show that OTFS casts the Doppler impairment as an additional source of diversity while avoiding the devastating effect of intercarrier interference.
4. Ultra-Reliable Low Latency Communication (URLLC). In this use case we show that OTFS exhibits resilience to narrowband interference thus allowing seamless co-existence with URLLC packets and other types of narrowband interference.
5. Potential use for mm-Wave communication. The discussion of this use case is a summary of a work in progress on the potential of OTFS for communication under high phase-noise impairment. In this context, we will explain how phase noise can be mitigated in the OTFS setting without sacrificing capacity.

# 2. OTFS PRINCIPLES

## 2.1. OTFS IN A NUTSHELL

OTFS is a modulation scheme that multiplexes QAM information symbols in a new signal representation called the delay-Doppler representation. In the mathematical literature, the delay-Doppler representation is sometimes referred to as the lattice representation of the Heisenberg group. The structure was later rediscovered by physicists who refer to it as the Zak representation[1]. The delay-Doppler representation generalizes time and frequency representations, rendering OTFS as a far reaching generalization of well known time and frequency modulation schemes such as TDM (Time Division Multiplexing, or single carrier multiplexing) where QAM symbols are multiplexed over consecutive time slots and FDM (Frequency Division Multiplexing, or multi-carrier multiplexing) where QAM symbols are multiplexed over consecutive frequency carriers. From a broader perspective, OTFS establishes a conceptual link between Radar and communication. These aspects of the theory are explained in Sections 2.2 and 2.3.

The OTFS waveforms optimally couple with the wireless channel in a way that captures the physics of the channel, yielding a high-resolution delay-Doppler Radar image of the constituent reflectors. This results in a simple symmetric coupling between the channel and the information carrying QAM symbols. The symmetry manifests itself through three fundamental properties:

- Invariance
- Separability
- Orthogonality

Invariance means that the coupling pattern is the same for all QAM symbols (i.e., all symbols experience the same channel or, put another way, the coupling is translation invariant). Separability (sometimes referred to as hardening) means that **all** the diversity paths are separated from one another and each QAM symbol experiences all the diversity paths of the channel. Finally, orthogonality means that the coupling is localized, which implies that the QAM symbols remain roughly orthogonal to one another at the receiver. The orthogonality property should be contrasted with conventional PN sequence-based CDMA modulations where every codeword introduces a global interference pattern that affects all the other codewords. The invariance property should be contrasted with TDM and FDM where the coupling pattern vary significantly among different time-frequency coherence intervals. This aspect of the theory is explained in Section 2.6

A variant of OTFS can be architected over an arbitrary multicarrier modulation scheme by means of a two-dimensional (symplectic) Fourier transform between a grid in the delay-Doppler plane and a grid in the reciprocal time-frequency plane. The Fourier relation gives rise to a family of orthogonal 2D basis functions on the time-frequency grid where each of these basis functions can be viewed as a codeword that spreads over multiple tones and multiple multicarrier symbols. This interpretation renders OTFS as a time-frequency spreading technique that generalizes CDMA. This aspect of the theory will be explained in Section 2.7.

[1] After Joshua Zak, Department of Physics, Technion – Israel Institute of Technology

## 2.2. THE DELAY-DOPPLER SIGNAL REPRESENTATION

To understand OTFS from first principles one should revisit the foundations of signal processing which at its core revolves around two basic **signal representations**. One is the time representation, where a signal is realized as a function of time (superposition of delta functions) and the other is the frequency representation where a signal is realized as a function of frequency (superposition of complex exponentials). These two representations are interchangeable using the Fourier transform.

The time and frequency representations are complementary to one another. The mathematical expression of this complementarity is captured by the Heisenberg uncertainty principle which states that a signal cannot be simultaneously localized to any desired degree in time and in frequency. Specifically, if a signal is localized in time then it is non localized in frequency and, reciprocally, if a signal is localized in frequency then it is non localized in time, as shown in Figure 1. This mathematical fact hides a deeper truth. As it turns out, there exists signals which **behave** as if they are simultaneously localized to any desired degree both in time and in frequency, a property which renders them optimal both for delay-Doppler Radar multi-target detection and for wireless communication (two use cases which turn out to be strongly linked). These special signals are naturally associated with localized pulses in a representation called the delay-Doppler representation. Signals in the delay-Doppler representation are special type of functions on a two-dimensional domain called the delay-Doppler plane whose points are parametrized by two variables $(\tau, \upsilon)$ where the first variable is called delay and the second variable is called Doppler.

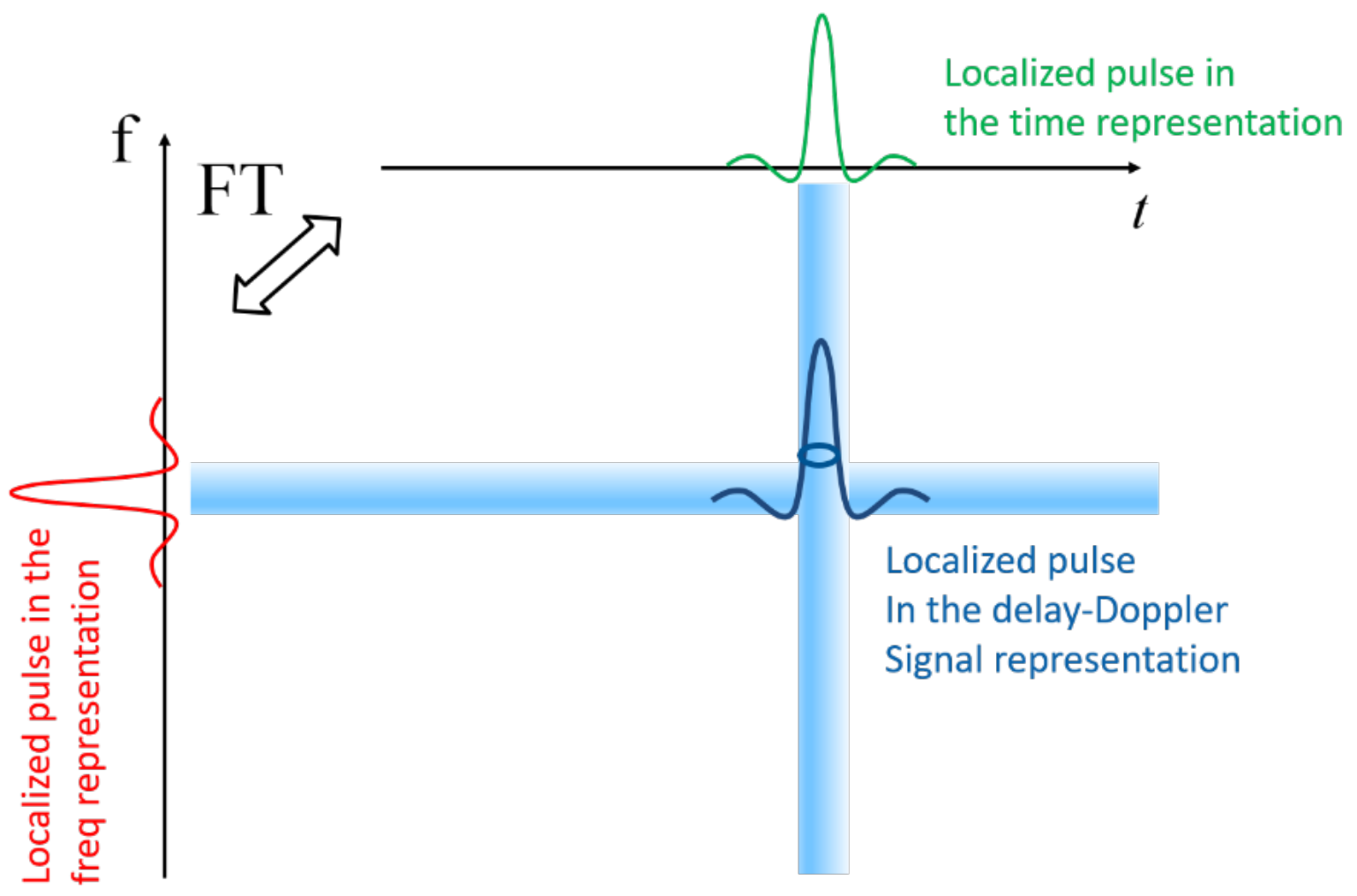

*Figure 1. Complementarity of time and frequency representations*

The delay-Doppler variables are commonly used in Radar and communication theory. In Radar, they are used to represent and separate moving targets by means of their delay (range) and Doppler (velocity) characteristics. In communication, they are used to represent channels by means of a superposition of time and frequency shift operations. The delay-Doppler channel representation is particularly meaningful in wireless communication, where it coincides with the delay-Doppler Radar image of the constituent reflectors [4]. Figure 2, shows an example of the delay-Doppler representation of a specific channel which is composed of two main reflectors which share similar delay (range) but differ in their Doppler characteristic (velocities).

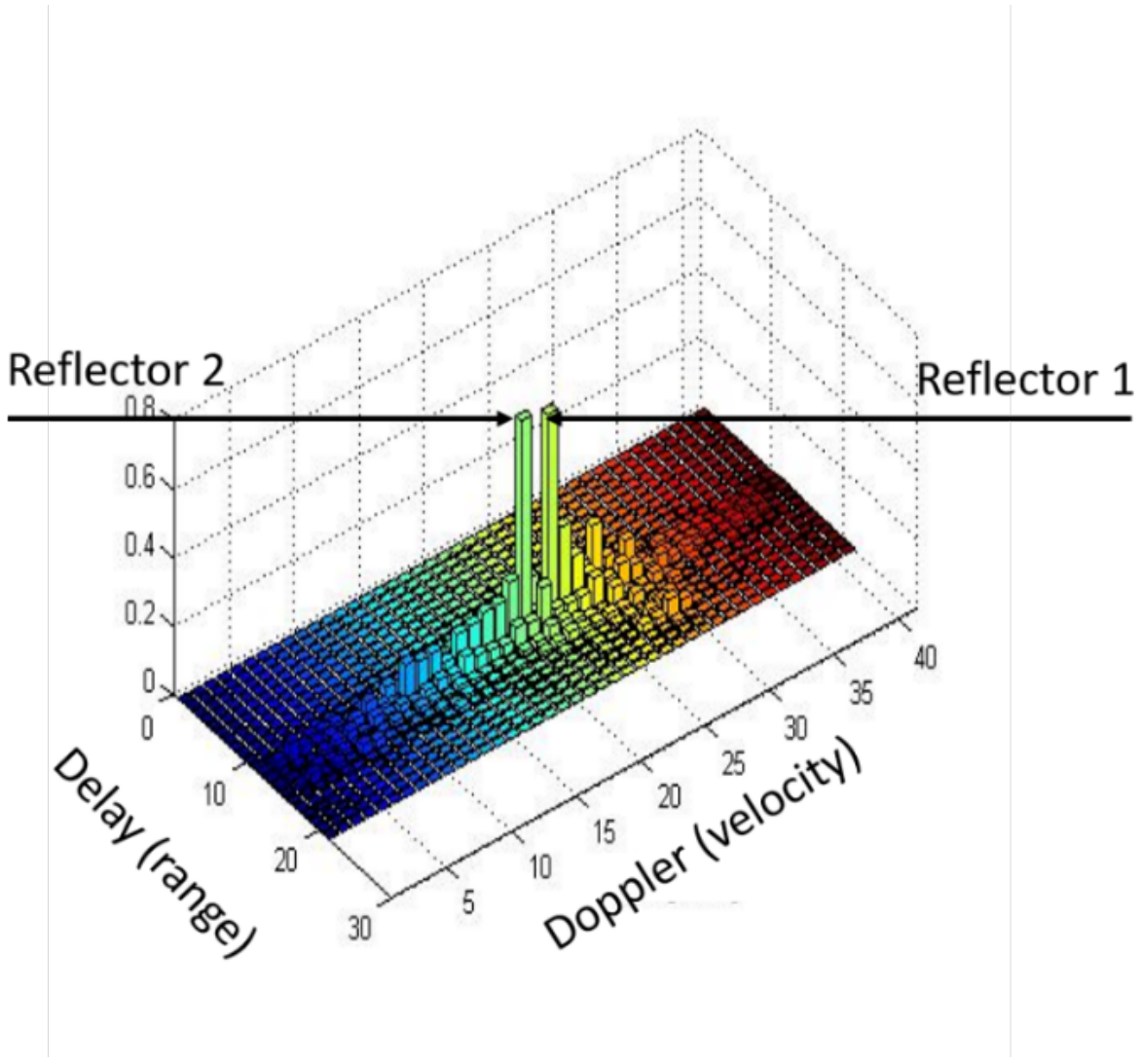

*Figure 2. The Delay-Doppler Impulse Response*

The use of the delay-Doppler variables to represent channels is well known. Less known is the fact that these variables can also be used to represent information-carrying signals in a way that is harmonious with the delay-Doppler representation of the channel. The delay-Doppler signal representation is mathematically subtler and requires the introduction of a new class of functions called quasi-periodic functions. To this end, we choose a delay period $\tau_r$ and a Doppler period $\nu_r$ satisfying the condition $\tau_r \cdot \upsilon_r = 1$ and thus defining a box of unit area, as shown in Figure 3. A delay-Doppler signal is a function $\phi(\tau, \upsilon)$ that satisfies the following quasi-periodicity condition:

$$\phi(\tau + n\tau_r, \upsilon + m\upsilon_r) = e^{j2\pi(n\upsilon\tau_r - m\tau\nu_r)}\phi(\tau, \upsilon)$$

which means that the function is periodic up to a multiplicative phase, i.e., the value of the function acquires a phase factor equal to $e^{j2\pi\nu\tau_r}$ for every traversal of delay period $\tau_r$ and, reciprocally, acquires a phase factor equal to $e^{-j2\pi\tau\nu_r}$ for every traversal of Doppler period $\nu_r$.

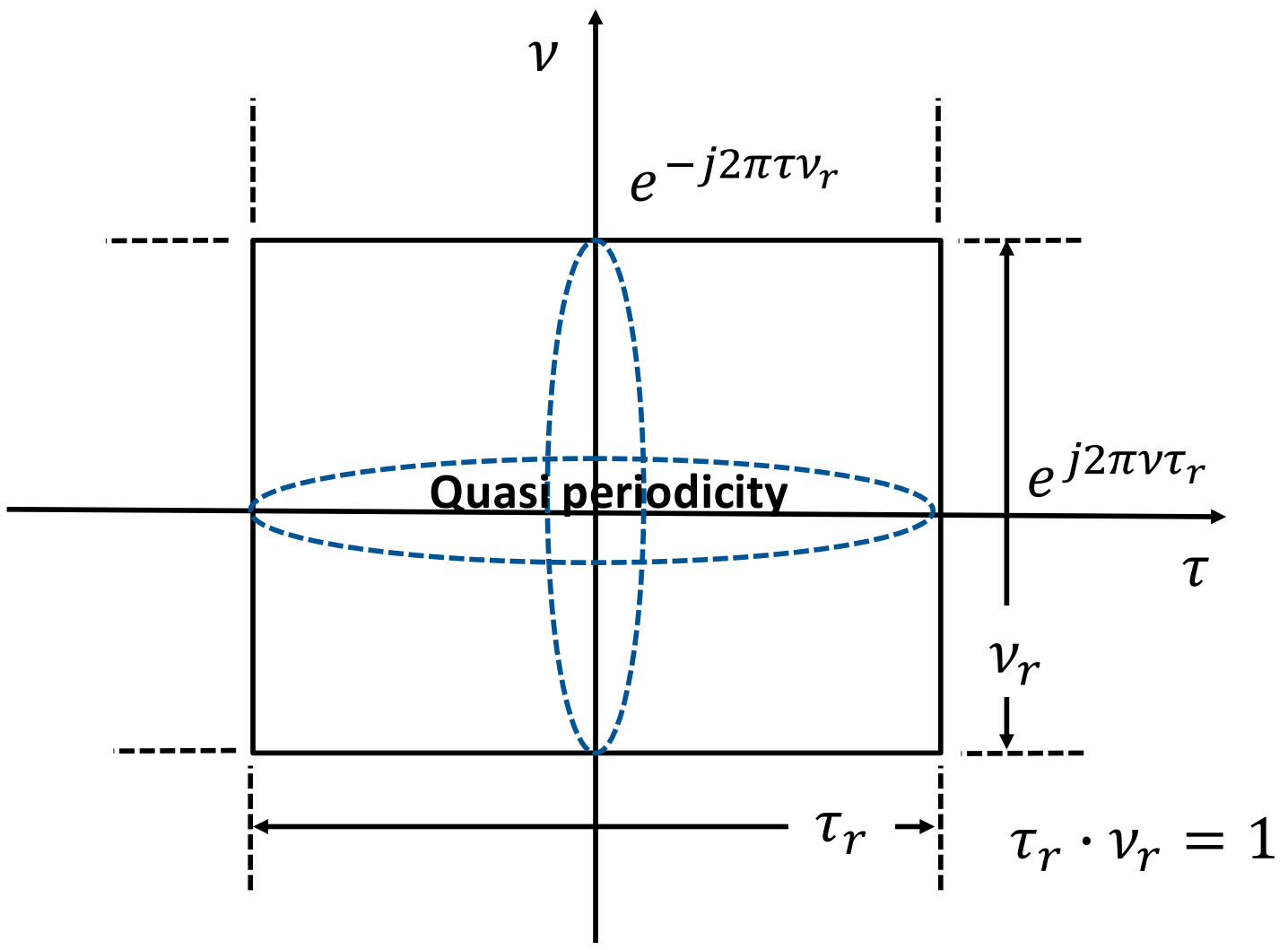

*Figure 3. Delay-Doppler quasi periodicity*

To summarize, there are three fundamental ways to represent a signal. The first way is as a function of time, the second way is as a function of frequency and the third way is as a quasi-periodic function of delay and Doppler. These three alternative representations are interchangeable by means of canonical transforms, as shown in Figure 4. The conversion between the time and frequency representations is carried through the Fourier transform. The conversion between the delay-Doppler and the time and frequency representations is carried by the Zak transforms $Z_t$ and $Z_f$ respectively ( [5], [6], [7] and [8]). The Zak transforms are realized by means of periodic Fourier integration formulas:

$$Z_t(\phi) = \int_0^{\upsilon_r} e^{j2\pi t\upsilon}\phi(t,\upsilon)d\upsilon$$

$$Z_f(\phi) = \int_0^{\tau_r} e^{-j2\pi t\upsilon}\phi(\tau,f)d\tau$$

In words, the Zak transform to the time representation is given by the inverse Fourier transform along a Doppler period and reciprocally, the Zak transform to the frequency representation is given by the Fourier transform along a delay period. We note that the quasi-periodicity condition is necessary to for the Zak transform to be a one-to-one equivalence between functions on the one-dimensional line and functions on the two-dimensional plane. Without it, a signal on the line will admit infinitely many delay-Doppler representations[2].

[2] In a sense the situation resembles the Fourier equivalence between sampled functions on the line and periodic functions on the line. Without imposing periodicity, a sampled function will have infinitely many representations in the Fourier domain.

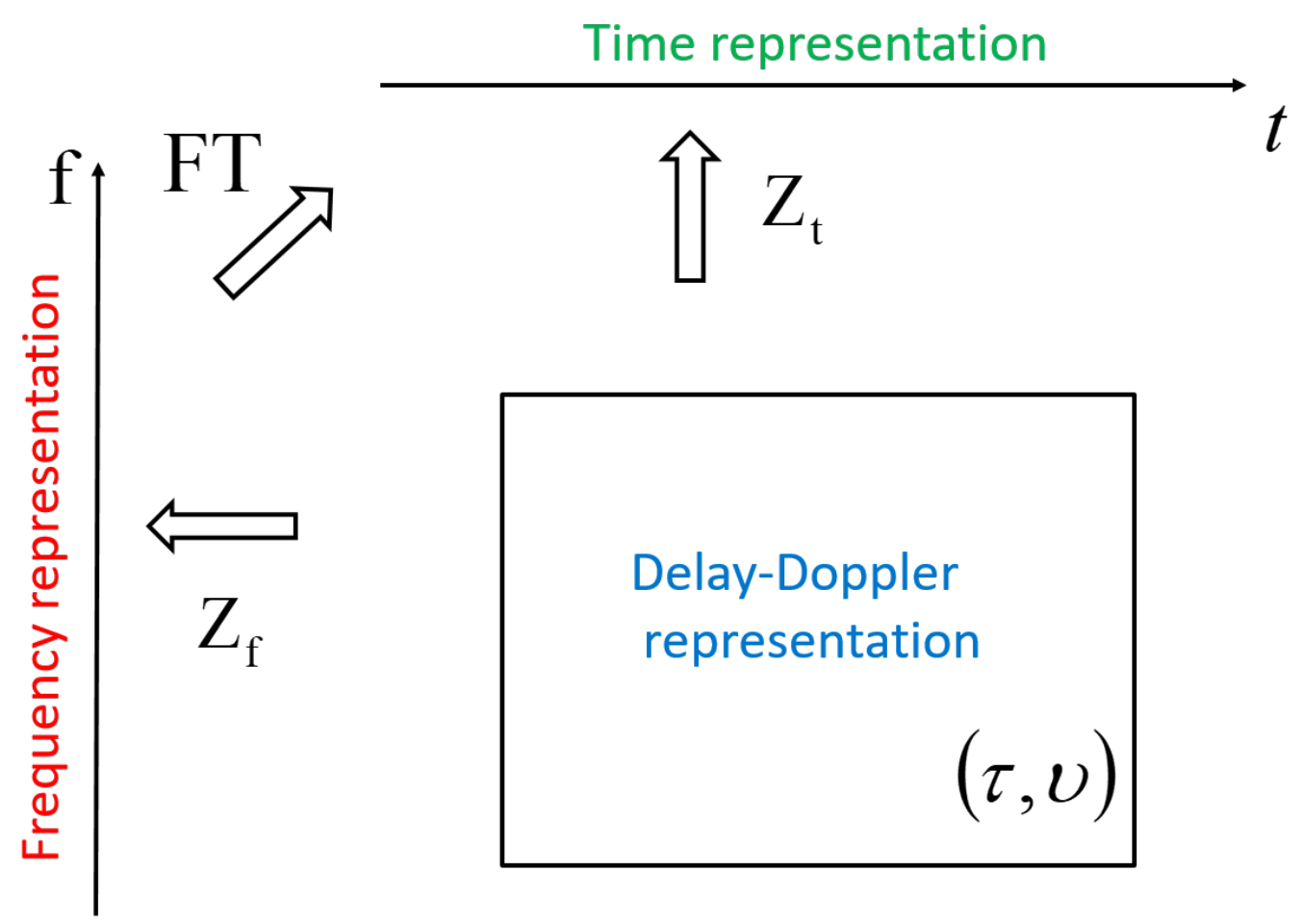

*Figure 4. The delay-Doppler representation*

## 2.3. THE GENERAL FRAMEWORK OF SIGNAL PROCESSING

The general framework of signal processing consists of three signal representations – (1) time, (2) frequency, and (3) delay-Doppler, interchangeable by means of canonical transforms. The setting can be neatly organized in a form of a triangle, as shown in Figure 5. The nodes of the triangle represent the three representations and the edges represent the canonical transformation rules converting between them.

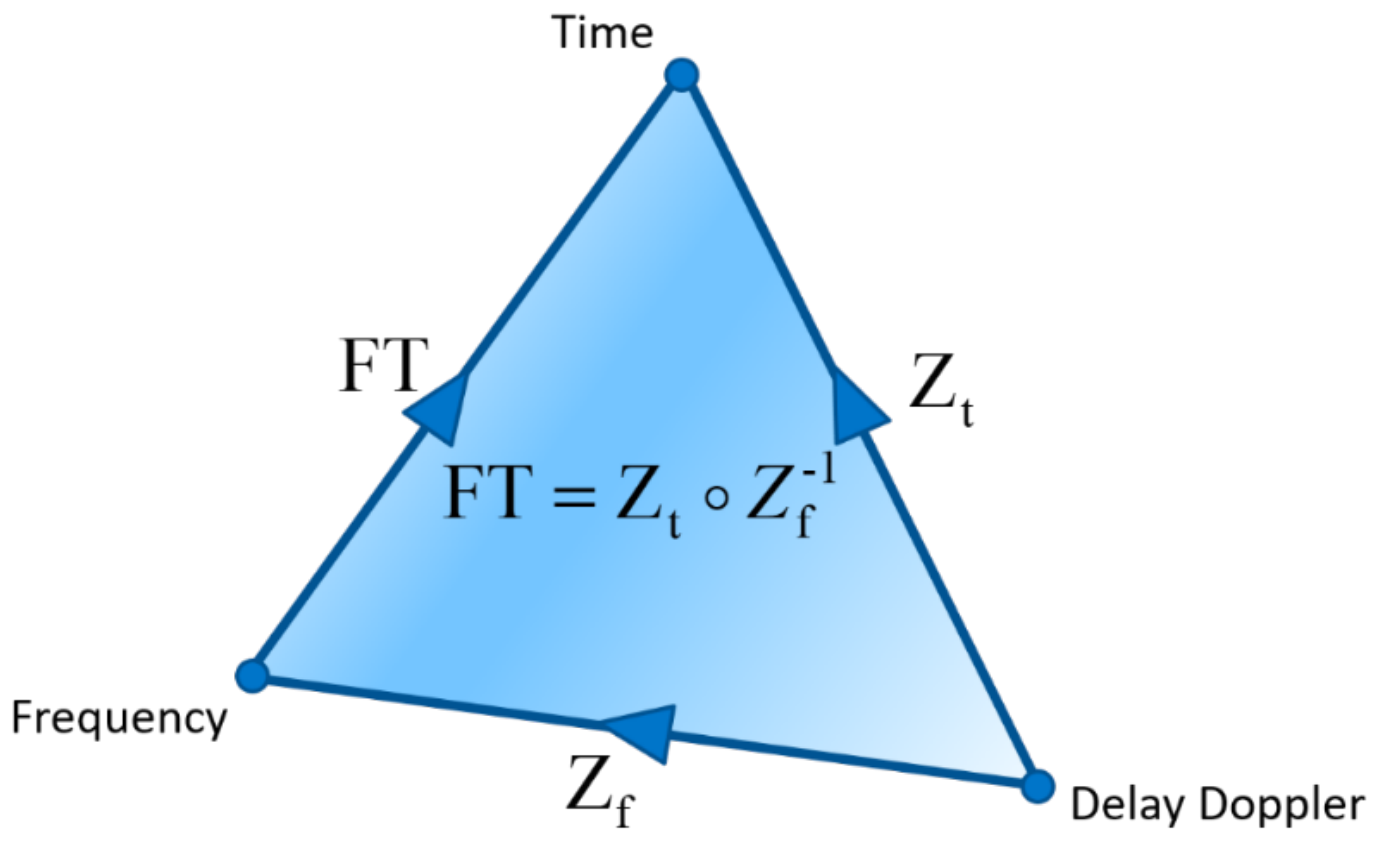

*Figure 5. The fundamental triangle*

An important property of this diagram is that the composition of any pair of transforms is equal to the remaining third one. In other words, traversing along the edges of the triangle results in the same answer no matter of which path is chosen. In particular, one can write the Fourier transform as a composition of two Zak transforms:

$$FT = Z_t \circ Z_f^{-1}$$

This means that instead of transforming from frequency to time using the Fourier transform one can alternatively transform from frequency to delay-Doppler using the inverse Zak transform $Z_f^{-1}$ and then from delay-Doppler to time using the Zak transform $Z_t$. The above decomposition yields an alternative algorithm for computing the Fourier transform which turns out to coincide with the fast Fourier transform algorithm discovered by Cooley-Tukey[3]. This striking fact is an evidence that the delay-Doppler representation silently plays an important role in classical signal processing.

Going up one level of abstraction, we note that the delay-Doppler representation is not unique but depends on a choice of a pair of periods $(\tau_r, \upsilon_r)$ satisfying the relation $\tau_r \cdot \upsilon_r = 1$. This implies that there is a continuous family of delay-Doppler representations, corresponding to points on the hyperbola $\upsilon_r = 1/\tau_r$, as shown in Figure 6. It is interesting to study what happens in the limits when the variable $\tau_r \rightarrow \infty$ and when the variable $\upsilon_r \rightarrow \infty$. In the first limit the delay period is extended at the expense of the Doppler period contracting, thus converging in the limit to a one-dimensional representation coinciding with the time representation. Reciprocally, in the second limit, the Doppler period is extended at the expense of the delay period contracting, thus converging in the limit to a one-dimensional representation coinciding with the frequency representation. Hence, the time and frequency representations can be viewed as limiting cases of the more general family of delay-Doppler representations.

All delay-Doppler representations are interchangeable by means of appropriately defined Zak transforms which satisfy commutativity relations generalizing the triangle relation discussed beforehand. This means that the conversion between any pair of representations along the curve is independent of which polygonal path is chosen to connect between them. On a philosophical note, the delay-Doppler representations and the associated Zak transforms constitute the primitive building blocks of signal processing giving rise, in particular, to the classical notions of time and frequency and the associated Fourier transformation rule.

[3] More accurately, the FFT algorithm amounts to a decomposition of the Fourier transform into a sequence of intermediate Zak transforms converting between the points of a polygonal decomposition of the delay-Doppler curve, explained below.

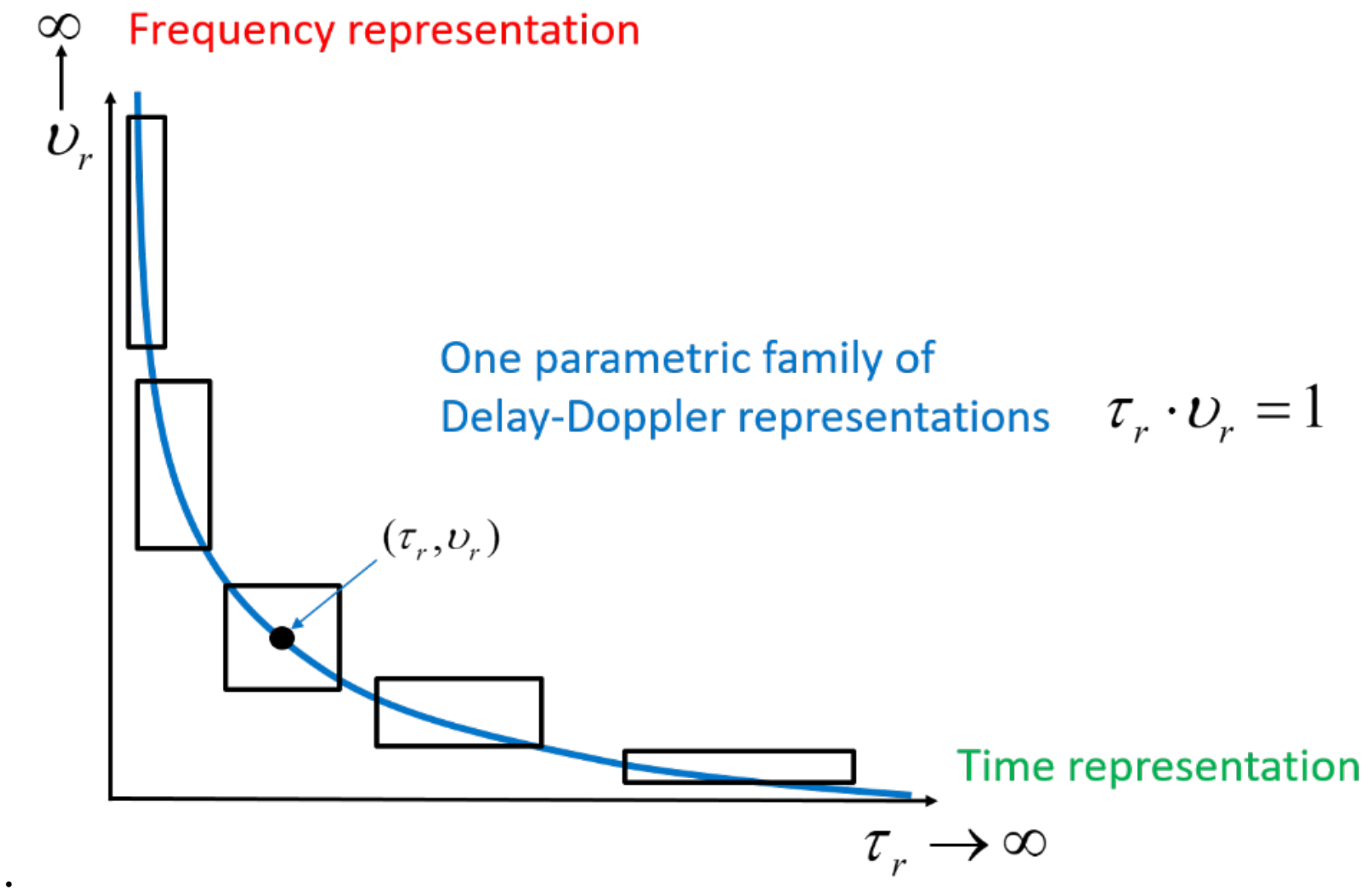

*Figure 6. Delay-Doppler parametric representation*

## 2.4. OTFS MODULATION SCHEME

Communication theory is concerned with transferring information through various physical media such as wired and wireless. The vehicle that couples a sequence of information-carrying QAM symbols with the communication channel is referred to as a modulation scheme. The channel-symbol coupling thus depends both on the physics of the channel and on the modulation carrier waveforms. Consequently, every modulation scheme gives rise to a unique coupling pattern which reflects the way the modulation waveforms interact with the channel.

Classical communication theory revolves around two basic modulation schemes which are naturally associated with the time and frequency signal representations. The first scheme multiplexes QAM symbols over localized pulses in the time representation and it is called TDM (Time Division Multiplexing). The second scheme multiplexes QAM symbols over localized pulses in the frequency representation (and transmits them using the Fourier transform) and it is called FDM (Frequency Division Multiplexing).

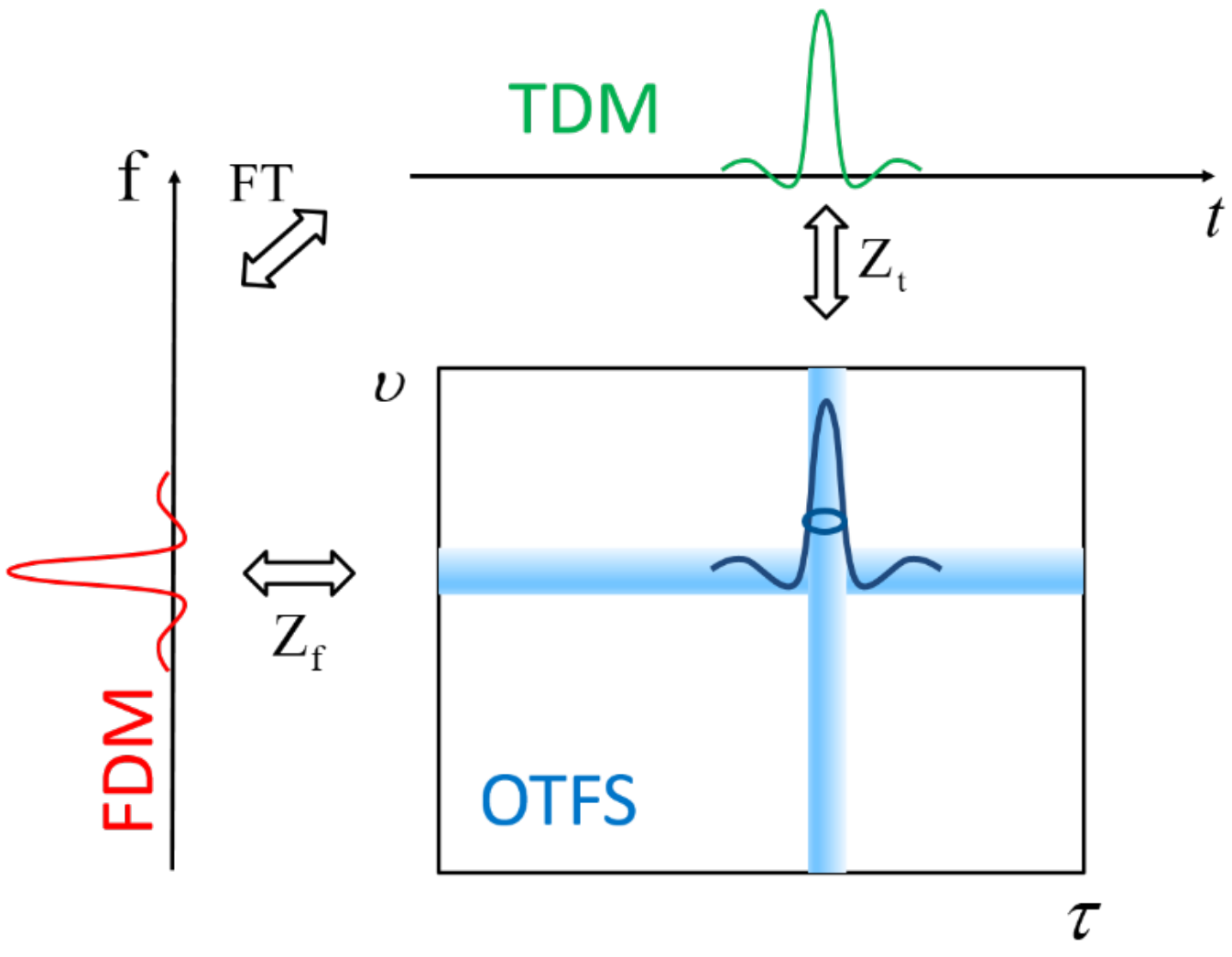

*Figure 7. Delay-Doppler modulation scheme*

It is interesting to convert the TDM and FDM carrier pulses to the delay-Doppler representation using the respective inverse Zak transforms. Converting the TDM pulse reveals a quasi-periodic function that is localized in delay but non localized in Doppler. Converting the FDM pulse reveals a quasi-periodic signal that is localized in Doppler but non localized in delay. The polarized non-symmetric delay-Doppler representation of the TDM and FDM pulses suggests a superior modulation based on symmetrically localized signals in the delay-Doppler representation, as shown in Figure 7. This new modulation scheme is referred to as OTFS, which stands for Orthogonal Time Frequency and Space.

There is an infinite family of OTFS modulation schemes corresponding to different delay-Doppler representations parameterized by points of the delay-Doppler curve (as shown in Figure 6). The classical time and frequency modulation schemes, TDM and FDM, appear as limiting cases of the OTFS family, when the delay and Doppler periods approach infinity, respectively. The OTFS family of modulation schemes smoothly interpolate between time and frequency division multiplexing.

## 2.5. THE OTFS CARRIER WAVEFORM

Up to this point we have used the abstract language of domains, signal representations and transforms to describe OTFS. In this section, we give an explicit description of the OTFS carrier waveform as a function of time. To this end, we choose a two-dimensional grid in the delay-Doppler plane specified by the following parameters:

$$\Delta\tau = \frac{\tau_r}{N}$$

$$\Delta\upsilon = \frac{\upsilon_r}{M}$$

The grid defined in this way consists of $N$ points along the delay period, with spacing $\Delta\tau$ and $M$ points along the Doppler period, with spacing $\Delta\upsilon$, resulting with a total of $NM$ grid points inside the fundamental rectangular domain. Next, we position a localized pulse, $w_{n,m}$, in the delay-Doppler representation at a specific grid point $(n\Delta\tau, m\Delta\upsilon)$. We note that the pulse is only localized inside the boundaries of the fundamental domain (enclosed by the delay-Doppler periods) and repeats itself quasi-periodically over the whole delay-Doppler plane, as shown in Figure 8 with $n = 3$ and $m = 2$. We assume $w_{n,m}$ is a product of two one-dimensional pulses:

$$w_{n,m}(\tau, \upsilon) = w_\tau(\tau - n\Delta\tau) \cdot w_\upsilon(\upsilon - m\Delta\upsilon)$$

where the first factor is localized along delay (time) and the second factor is localized along Doppler (frequency). In a sense, the delay-Doppler two-dimensional pulse is a stitching of the one-dimensional TDMA and OFDM pulses. To describe the structure of $w_{n,m}$ in the time representation we need to compute the Zak transform:

$$Z_t(w_{n,m})$$

A direct verification using the formula of the Zak transform reveals that the resulting waveform is a pulse train shifted in time and in frequency, where the shift in time is equal to the delay coordinate $n\Delta\tau$ and the shift in frequency is equal to the Doppler coordinate $m\Delta\upsilon$. Locally, the shape of each pulse is related to the delay pulse, $w_\tau$, and, globally, the shape of the total train is related to the Fourier transform of the Doppler pulse, $w_\upsilon$. Moving the grid point along delay causes each pulse in the train to shift along time by the same displacement, resembling TDM. Reciprocally, moving the grid point along Doppler causes a shift in frequency of the whole train by the same frequency displacement, resembling OFDM. In other words, the local structure of the OTFS carrier waveform resembles that of TDM while the global structure resembles that of FDM.

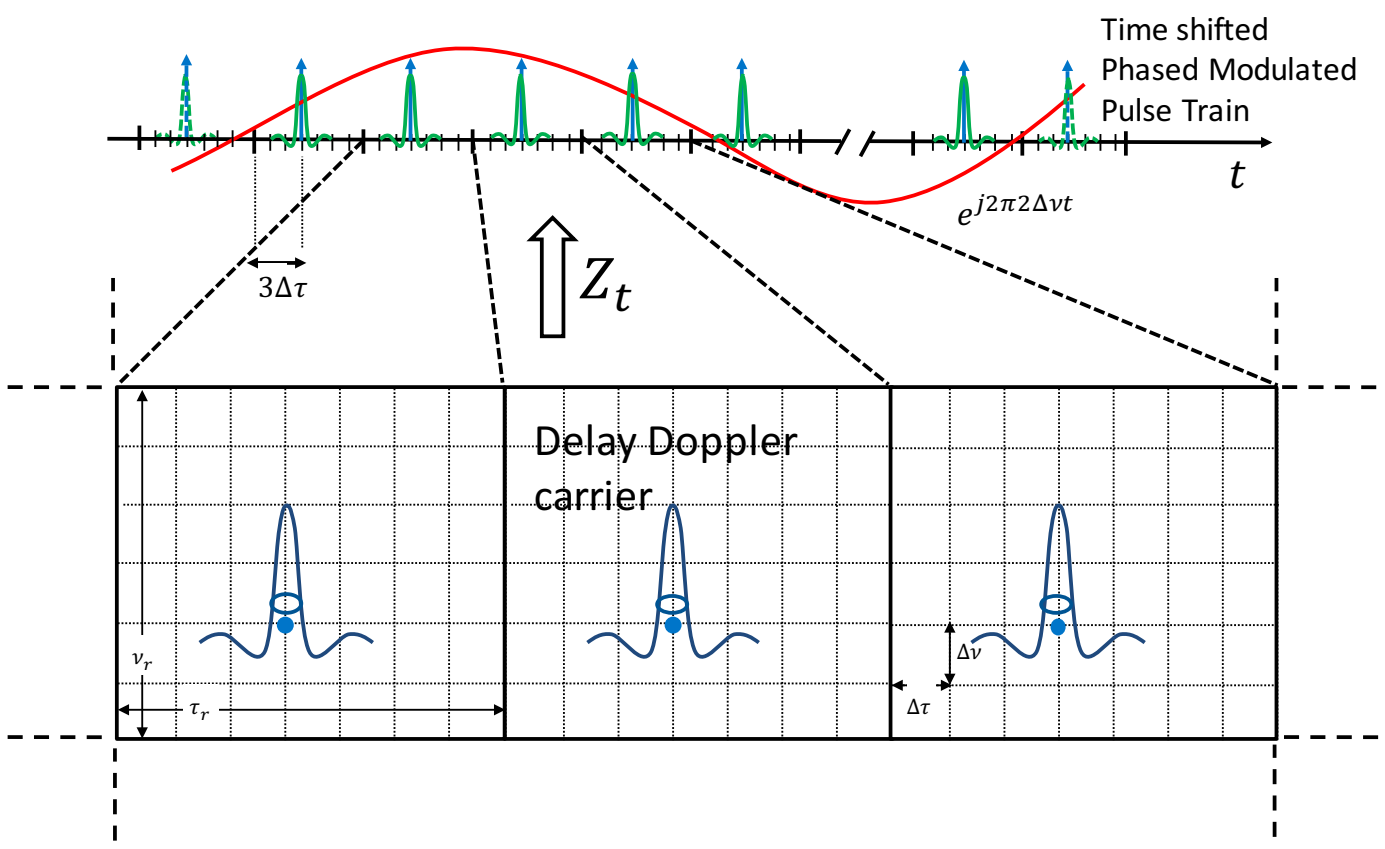

*Figure 8. The OTFS carrier waveform*

## 2.6. THE DELAY-DOPPLER CHANNEL SYMBOL COUPLING

The wireless channel is governed by simple physics. It is composed of a collection of specular reflectors, some of which are static and some of which are moving. The transmitted waveform propagates through the medium and bounces off each reflector. The signal that arrives at the receiver is a superposition of the direct signal and the reflected echoes. Each of the reflected echoes arrives at the receiver at a delayed time (multipath effect) and possibly also shifted in frequency (Doppler effect) due to the relative velocity between the reflector and the transmitter/receiver. The channel physics is mathematically modeled through the delay-Doppler impulse response where each tap represents a cluster of reflectors with specific delay and Doppler characteristics, as shown in Figure 4. Our goal in this section is to describe the channel-symbol coupling (CSC for short) between the wireless channel and the OTFS carrier waveform given by a localized pulse in the delay-Doppler representation. As a motivation, we first discuss the channel-symbol coupling of the TDM and FDM pulses.

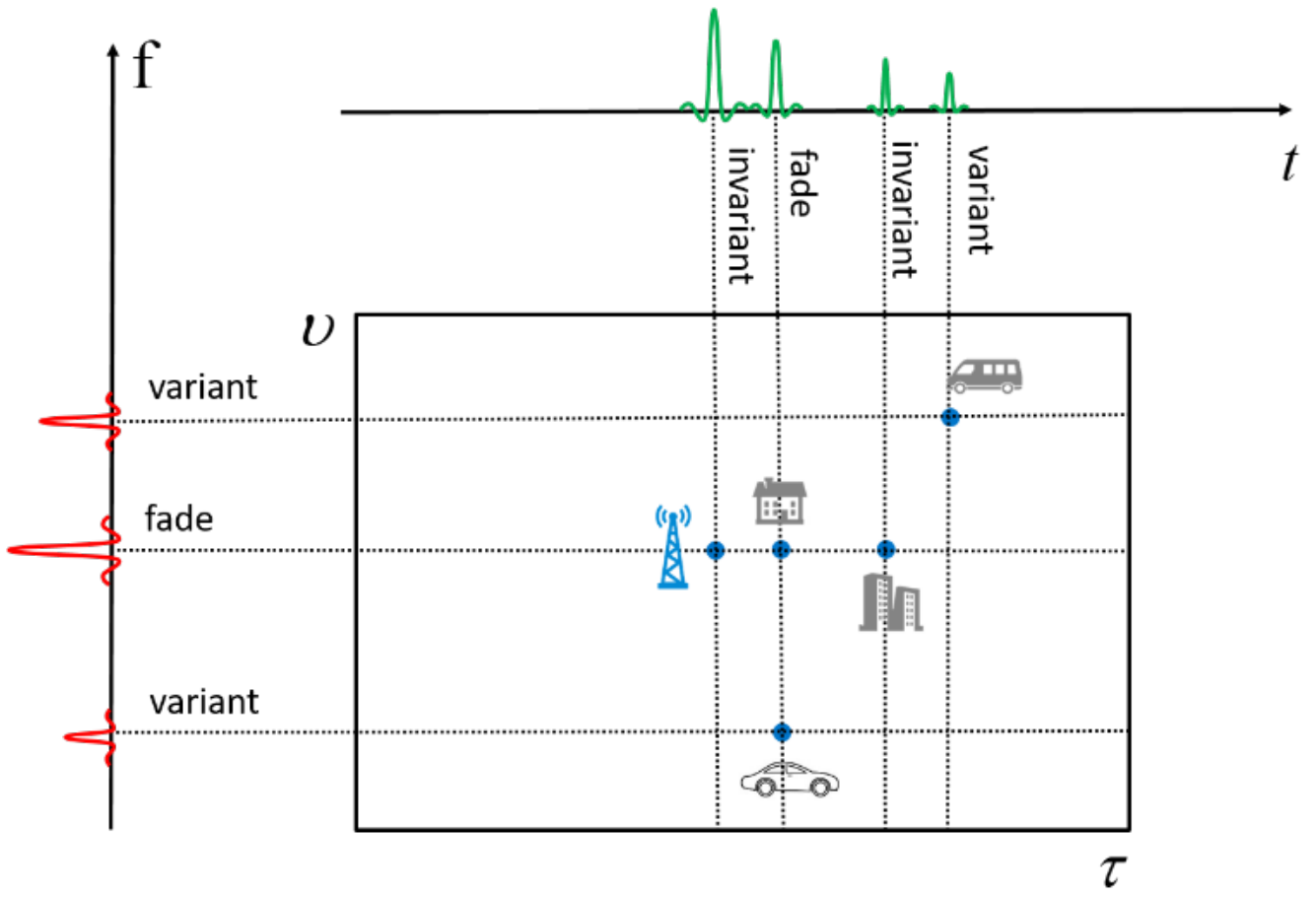

*Figure 9. TDM and FDM channel-symbol couplings*

Transmitting a localized TDM pulse in the time representation gives rise at the receiver to a configuration of echoes which appear at specific time displacements which corresponds to the multipath delays imposed by the various reflectors. The phase and amplitude of each echo depend on the initial position of the transmitted pulse and might change significantly among different coherence time intervals – a phenomenon referred to as time selectivity. There are two mechanisms involved. The phase of the echo changes due to the Doppler effect and the amplitude of the echo changes due to destructive superposition of numerous reflectors sharing the same delay but differing in Doppler, resulting from the inability of the TDM pulse to separate reflectors along Doppler. In Figure 9, counting the TDM echoes from left to right we see that the first and third echoes are due to static reflectors hence are time invariant, the fourth echo is due to moving reflector thus is time varying and the second echo is due to superposition of two reflectors, one of which is moving thus is fading.

Reciprocally, transmitting a localized FDM pulse in the frequency representation gives rise at the receiver to a configuration of echoes at specific frequency displacements which correspond to the Doppler shifts induced by the various reflectors. The phase and amplitude of each echo depends on the initial position of the transmitted pulse and might change significantly among different coherence frequency intervals – a phenomenon referred to as frequency selectivity. The phase of the echo changes due to the multipath effect and the amplitude of the echo changes due to destructive superposition of numerous reflectors sharing the same Doppler but perhaps differing in delay, resulting from the inability of the FDM pulse to separate reflectors along delay. For example, in Figure 9, counting the received FDM echoes from bottom up, we see that the first and third echoes are frequency varying and the second echo is due to superposition of three static reflectors thus is fading.

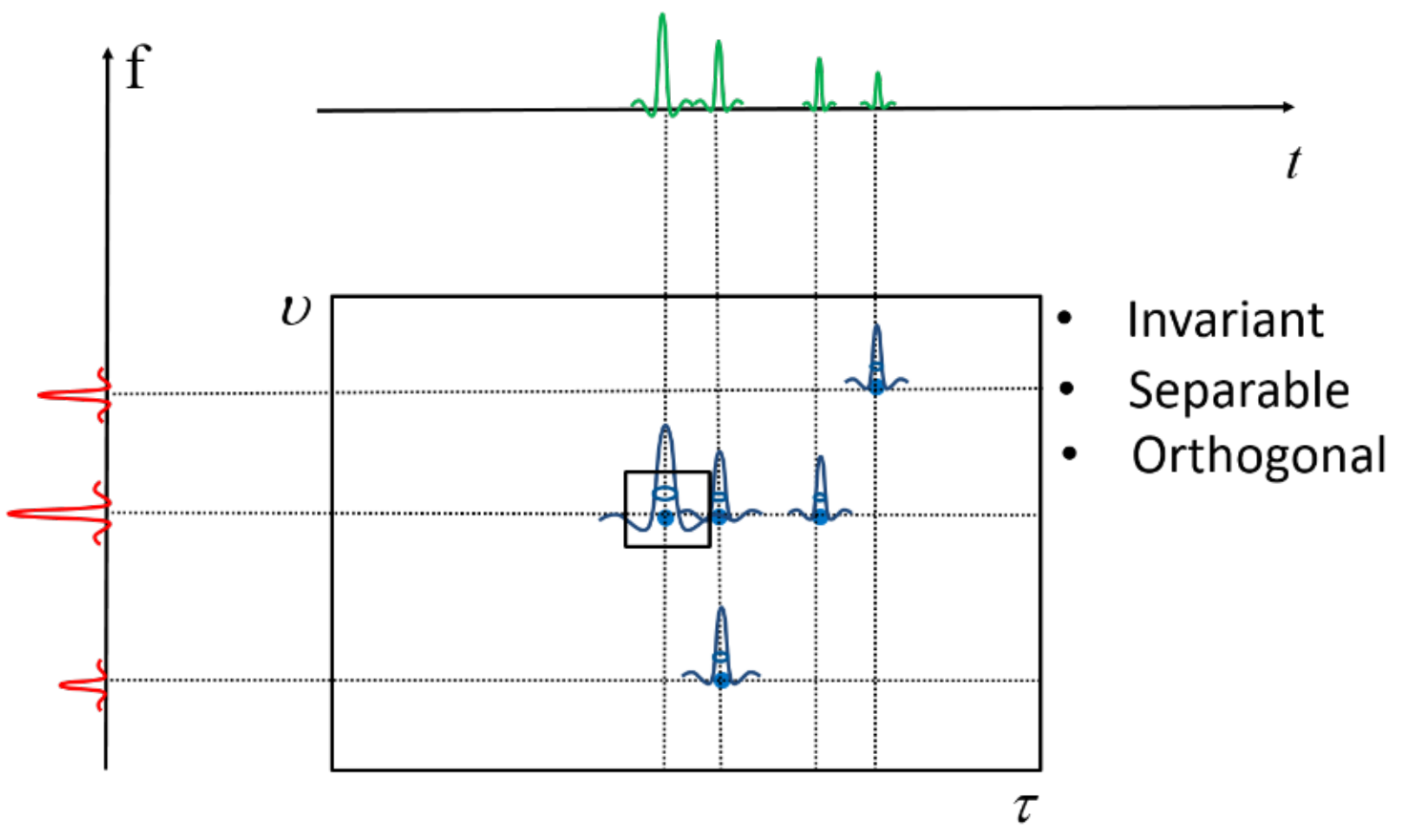

*Figure 10. Delay-Doppler channel-symbol coupling*

Transmitting a localized OTFS pulse in the delay-Doppler representation gives rise at the receiver to a configuration of echoes that appear at specific delay-Doppler displacements, which corresponds to the delay and Doppler shifts induced by the various reflectors, as shown in Figure 10. In contrast to the previous two cases, the following properties now hold:

- CSC **invariance**: the phase and amplitude of the delay-Doppler echoes are independent of the location of the original pulse inside the fundamental domain, since the delay and Doppler periods are smaller than the coherence time and bandwidth of the channel respectively.
- CSC **separability**: all the reflections are separated from one another along their delay and Doppler characteristics, hence their effects do not add up destructively and there is no loss of energy on the QAM symbol level.

- CSC **orthogonality**: the received echoes are confined to a small rectangular box around the transmitted pulse with dimensions equal the delay and Doppler spread of the channel which are much smaller than the outer delay and Doppler periods. As result, when two transmit pulses are geometrically separated at the transmitter they will remain orthogonal at the receiver.

An alternate way to express the OTFS channel-symbol coupling is as a two-dimensional convolution[4] between the delay-Doppler impulse response and the QAM symbols. This can be seen in Figure 11, which shows numerous delta functions (representing QAM symbols) convolved with the delay-Doppler impulse response of the channel.

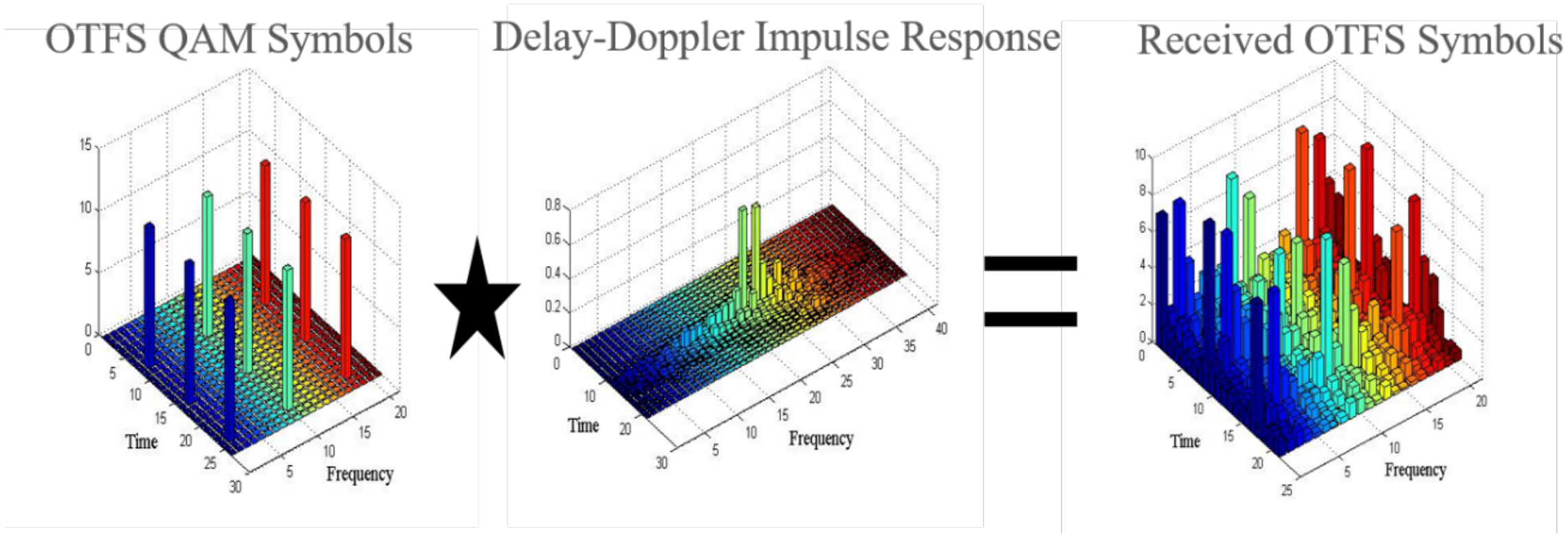

*Figure* 11. *2D Channel Convolution*

## 2.7. MULTICARRIER INTERPRETATION OF OTFS

In this section, we describe a variant of OTFS that is more adapted to the classical multicarrier formalism of time-frequency grids and filter-banks and illuminates aspects of OTFS that are not apparent from the Zak definition. One consequence of the new definition is that OTFS can be viewed as a time-frequency spreading scheme composed of a collection of two-dimensional basis-functions (or codewords) defined over a reciprocal time-frequency grid. Another consequence is that OTFS can be architected as a simple pre-processing step over an arbitrary multicarrier modulation such as OFDM. The new definition is based on Fourier duality relation between a grid in the delay-Doppler plane and a reciprocal grid in the time-frequency plane.

The delay-Doppler grid consists of $N$ points along delay with spacing $\Delta\tau = \tau_r/N$ and $M$ points along Doppler with spacing $\Delta\upsilon = \upsilon_r/M$ and the reciprocal time-frequency grid consists of $N$ points along frequency with spacing $\Delta f = 1/\tau_r$ and $M$ points along time with spacing $\Delta t = 1/\upsilon_r$. The two grids are shown in Figure 12. The parameter $\Delta t$ is the multicarrier symbol duration and the parameter $\Delta f$ is the subcarrier spacing. The time-frequency grid can be interpreted as a sequence of $M$ multicarrier symbols each consisting of $N$ tones or subcarriers. We note that the bandwidth of the transmission $B =$

[4] The precise mathematical description of the CSC is by means of operation called twisted convolution (also called Heisenberg convolution) of the delay-Doppler impulse response with the QAM symbols

$M\Delta f$ is inversely proportional to the delay resolution $\Delta\tau$ and the duration of the transmission $T = M\Delta t$ is inversely proportional to the Doppler resolution $\Delta\tau$.

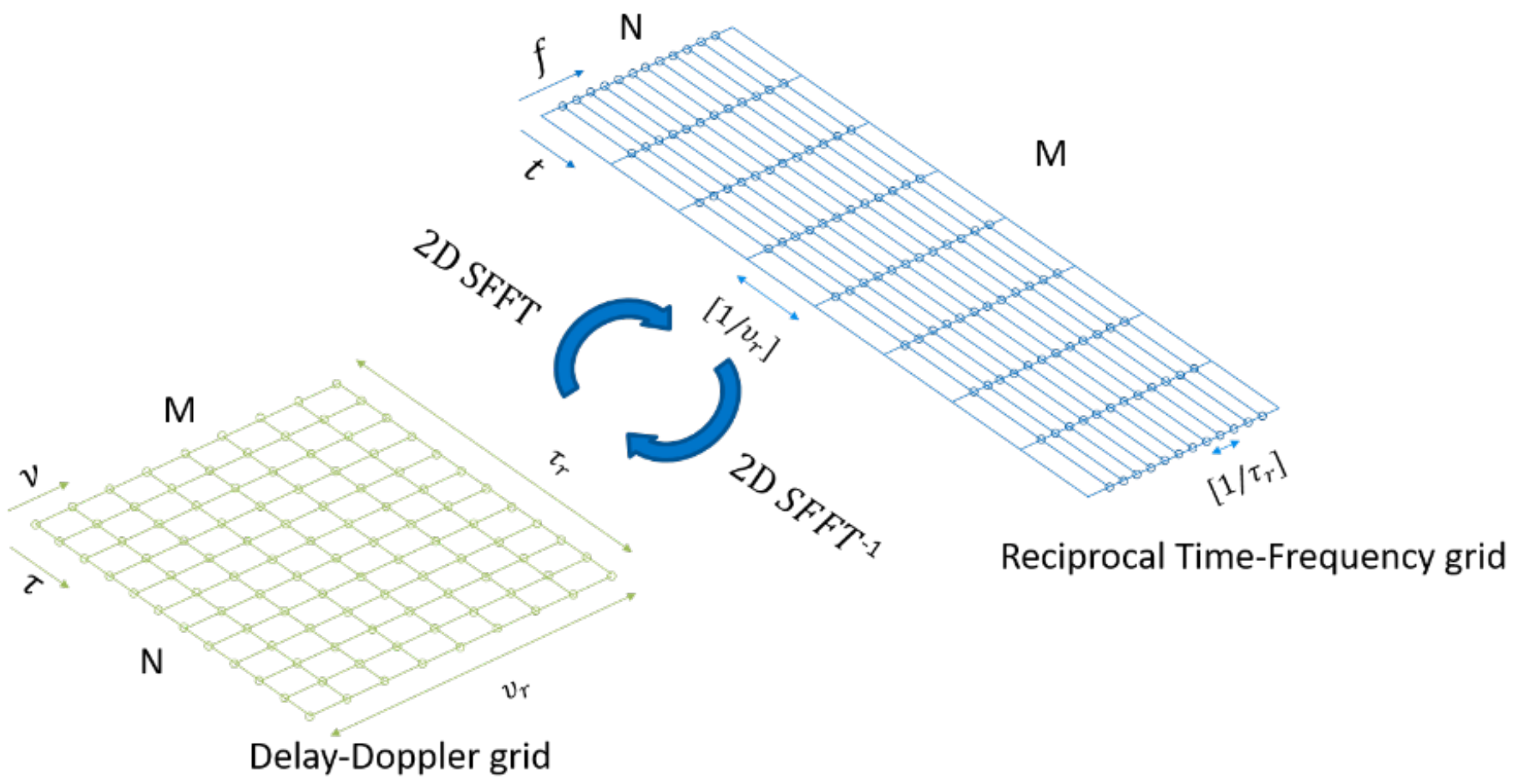

*Figure 12. Symplectic Fourier Duality*

The Fourier relation between the two grids is realized by a variant of the two-dimensional finite Fourier transform called the finite symplectic finite Fourier transform (SFFT for short). The SFFT sends an $N\times M$ delay-Doppler matrix $x(n\Delta\tau, m\Delta v)$ to a reciprocal $M\times N$ time-frequency $X(m'\Delta t, n'\Delta f)$ via the following summation formula:

$$X(m'\Delta t, n'\Delta f) = \sum_{n=0}^{N-1}\sum_{m=0}^{M-1} e^{j2\pi\left(\frac{m'm}{M}-\frac{n'n}{N}\right)} x(n\Delta\tau, m\Delta v)$$

where the term "symplectic" refers to the specific coupling form $m'm/M - n'n/N$ inside the exponent. One can easily verify that the SFFT transform is equivalent to an application of an $N$-dimensional FFT along the columns of the matrix $x(n,m)$ in conjunction with an $M$-dimensional IFFT along its rows.

The multicarrier interpretation of OTFS is the statement that the Zak transform of an $N\times M$ delay-Doppler matrix can be computed alternatively by first transforming the matrix to the time-frequency grid using the SFFT and then transforming the resulting reciprocal matrix to the time domain as a sequence of $M$ multicarrier symbols of size $N$ through a conventional multicarrier transmitter, i.e., IFFT transform of the columns. Hence, using the SFFT transform, the OTFS transceiver can be overlaid as a pre- and post-processing step over a multicarrier transceiver. The multicarrier transceiver of OTFS is depicted in Figure 13 along with a visual representation of the doubly selective multiplicative CSC in the time-frequency domain and the corresponding invariant convolutive delay-Doppler CSC.

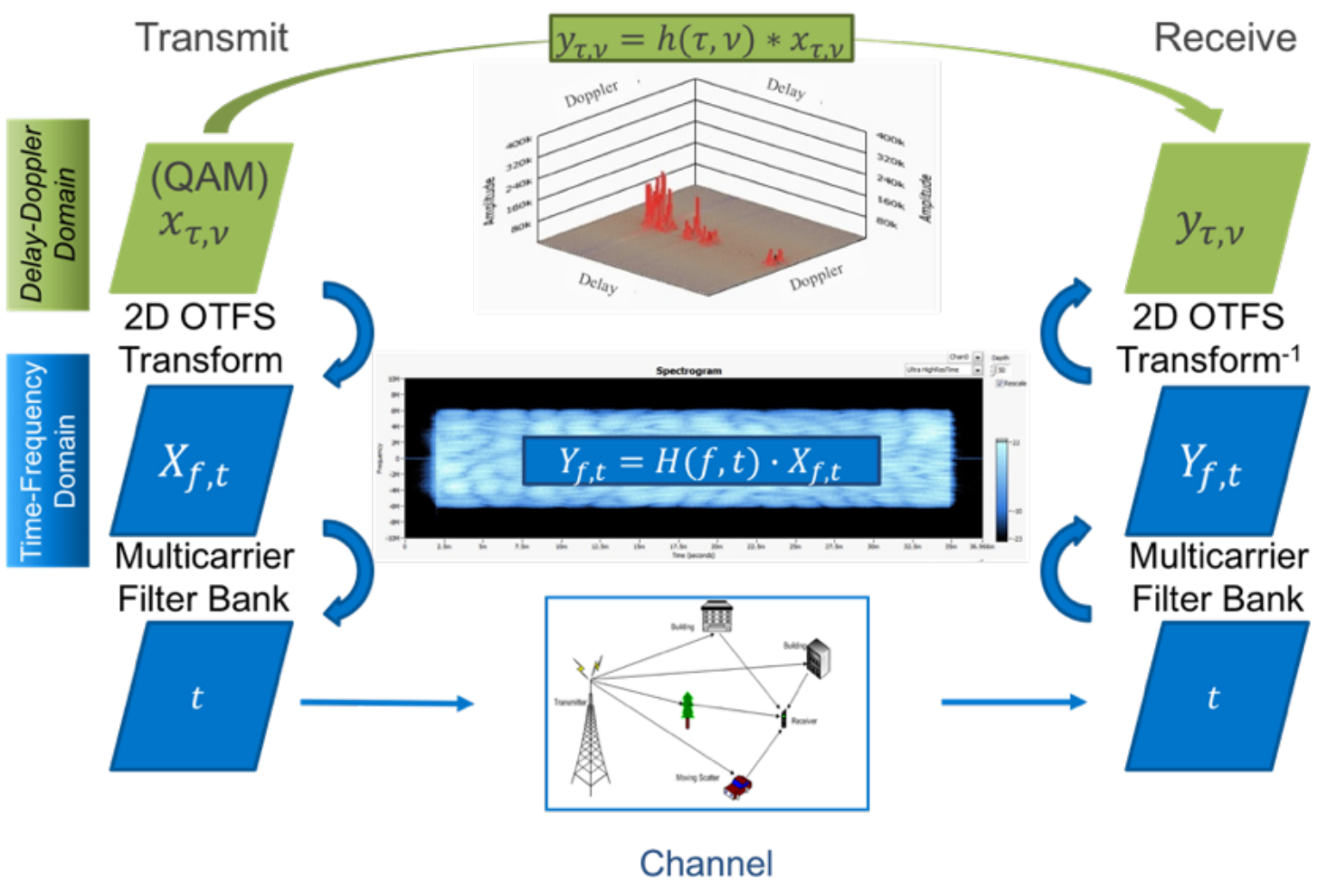

*Figure 13. Multicarrier OTFS Processing Steps*

The multicarrier interpretation casts OTFS as a time-frequency spreading technique where each delay-Doppler QAM symbol $x(n\Delta\tau, m\Delta\upsilon)$ is carried over a two-dimensional spreading 'code' or sequence on the time-frequency grid, given by the following symplectic exponential function:

$$\psi_{n,m}(m'\Delta t, n'\Delta f) = e^{j2\pi\left(\frac{mm'}{M} - \frac{nn'}{N}\right)}$$

where the slope of this function along time is given by the Doppler coordinate $m\Delta\upsilon$ and the slope along frequency is given by the delay coordinate $n\Delta\tau$ (see examples in Figure 14). Thus, the analogy to two dimensional CDMA is seen, where the codewords are 2D complex exponentials that are orthogonal to each other.

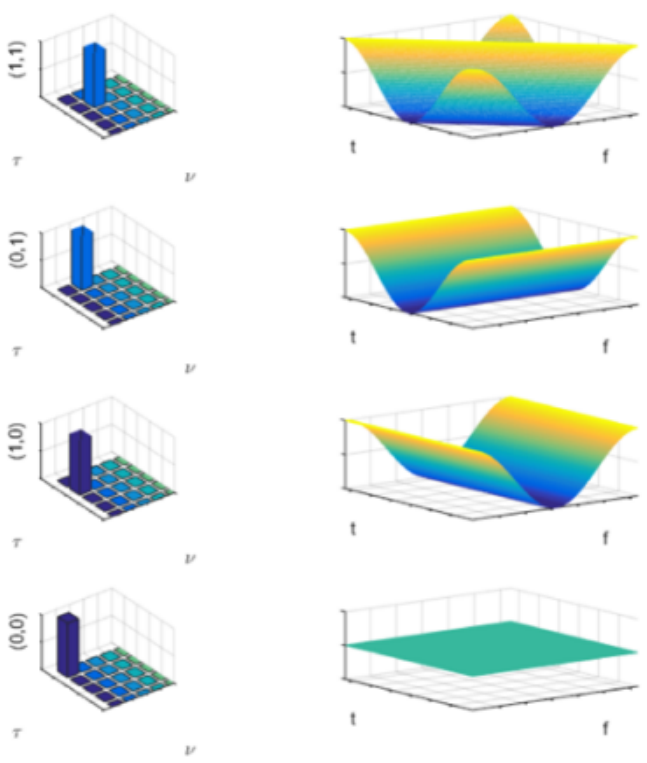

*Figure 14. OTFS Time-Frequency Basis Functions*

From a broader perspective, the Fourier duality relation between the delay-Doppler grid and the time-frequency grid establishes a mathematical link between Radar and communication where the first theory is concerned with maximizing the resolution of separation between reflectors/targets according to their delay-Doppler characteristics and the second is concerned with maximizing the amount of information that can be reliably transmitted through the communication channel composed of these reflectors.

# 3. DELAY-DOPPLER EQUALIZATION AND PRECODING

In this section, we discuss the principles of equalization and precoding when the QAM symbols are multiplexed in the delay-Doppler domain as in the case of OTFS and compare it with the situation when the QAM symbols are multiplexed in the time-frequency domain as in the case of multicarrier modulations. We focus on the context of multi-user MIMO (MU-MIMO for short) where a set of users are communicating simultaneously over the same bandwidth with a base-station equipped with multiple antennas.

## 3.1. EQUALIZATION

In the uplink direction, the streams from the different users arrive to the base-station when they are superimposed on one another and the base-station must separate between them by means of equalization. We assume $L_u$ users, each equipped with a single antenna, are transmitting to a base-station equipped with $L_b$ antennas. In the multicarrier setting the users are multiplexing their QAM symbols over a region of the time-frequency grid. Under these assumptions, the uplink channel is decoupled into a parallel (orthogonal) system of simple MIMO channels over the points of the time-frequency grid such that to every grid point $(m\Delta t, n\Delta f)$ there corresponds a local channel equation of the form:

$$Y_{m,n} = \boldsymbol{U}_{m,n} \cdot X_{m,n} + W_{m,n}$$

where $X_{m,n}$ is the vector of $L_u$ QAM symbols transmitted from the different users and $\boldsymbol{U}_{m,n}$ is an $L_b \times L_u$ matrix representing the local coupling between the user's and the base-station antennas. To retrieve the information of the users, the base-station must detect the QAM symbols composing the vector $X_{m,n}$. To maximize throughput, the QAM symbols must be detected jointly using a maximum likelihood sphere detector. The sphere detector is an iterative algorithm and its convergence rate critically depends on the condition number (the ratio between the maximum and minimum eigenvalues) of the auto-correlation matrix of the local channel:

$$\boldsymbol{R}_{m,n} = \boldsymbol{U}_{m,n}^{*}\boldsymbol{U}_{m,n}$$

where the superscript * denotes Hermitean transpose. When the condition number is high, the algorithm exhibits a critical slowdown which results in a complexity toll that is exponential in the MIMO order, i.e., the number of users. In the presence of channel time-frequency selectivity, a sizable portion of the auto-correlation matrices might exhibit high condition number and the resulting complexity toll becomes a formidable obstruction for scaling the system with the number of users.

There are two approaches to manage the receiver performance-complexity tradeoff. The first approach is to reduce the complexity of the detector by limiting the number of iterations at the expense of compromising performance. The second approach is to maintain performance by accelerating the convergence rate of the sphere detector using lattice reduction techniques at the expense of higher complexity due to the need to recalculate the reduced lattice basis for every coherence time and frequency interval. In other words, the time-frequency selectivity of the channel introduces a recalculation factor that results in a large complexity toll. We note that modern commercial MIMO systems typically implement the first approach. In practice, a full sphere detector is never used beyond the case of four spatial streams, due to the high complexity toll and most implementations use, instead, reduced-complexity variants with limited number of iterations.

The performance-complexity tradeoff of the receiver can be improved significantly by multiplexing the QAM symbols over the delay-Doppler grid where the channel-symbol coupling is invariant, separable and orthogonal (given by convolution with the delay-Doppler impulse response). For the sake of the explanation, we assume few simplified approximations. We assume the delay-Doppler uplink channel decouples into a parallel system of *identical* MIMO channels over the points of the delay-Doppler grid such that for each grid point $(n\Delta\tau, m\Delta\upsilon)$ there corresponds a local channel equation of the form:

$$y_{n,m} = \boldsymbol{u} \cdot x_{n,m} + w_{n,m}$$

where $x_{m,n}$ is the vector of QAM symbols transmitted from the different users and $\boldsymbol{u}$ is an $L_b \times L_u$ matrix representing the global coupling between the users and the base-station antennas. We further assume that the auto-correlation matrix $\boldsymbol{r} = \boldsymbol{u}^{*}\boldsymbol{u}$ is equal to the arithmetic average of all the local time-frequency auto-correlation matrices, that is:

$$\boldsymbol{r} = \frac{1}{NM}\sum_{n,m} \boldsymbol{R}_{m,n}$$

Since the time-frequency channel matrices, $\boldsymbol{U}_{m,n}$, are roughly independent from one another, the delay-Doppler matrix $\boldsymbol{r}$ has a lower condition number than a typical $\boldsymbol{R}_{m,n}$ due to averaging. The lower condition number of the matrix $\boldsymbol{r}$ implies faster convergence rate of the sphere detector thus rendering the detection problem over the delay-Doppler grid a much easier computational task. In addition, lattice reduction techniques can be efficiently used to accelerate the convergence rate even further since the reduced basis need to be computed only once per frame of $NM$ QAM symbols, due to invariance.

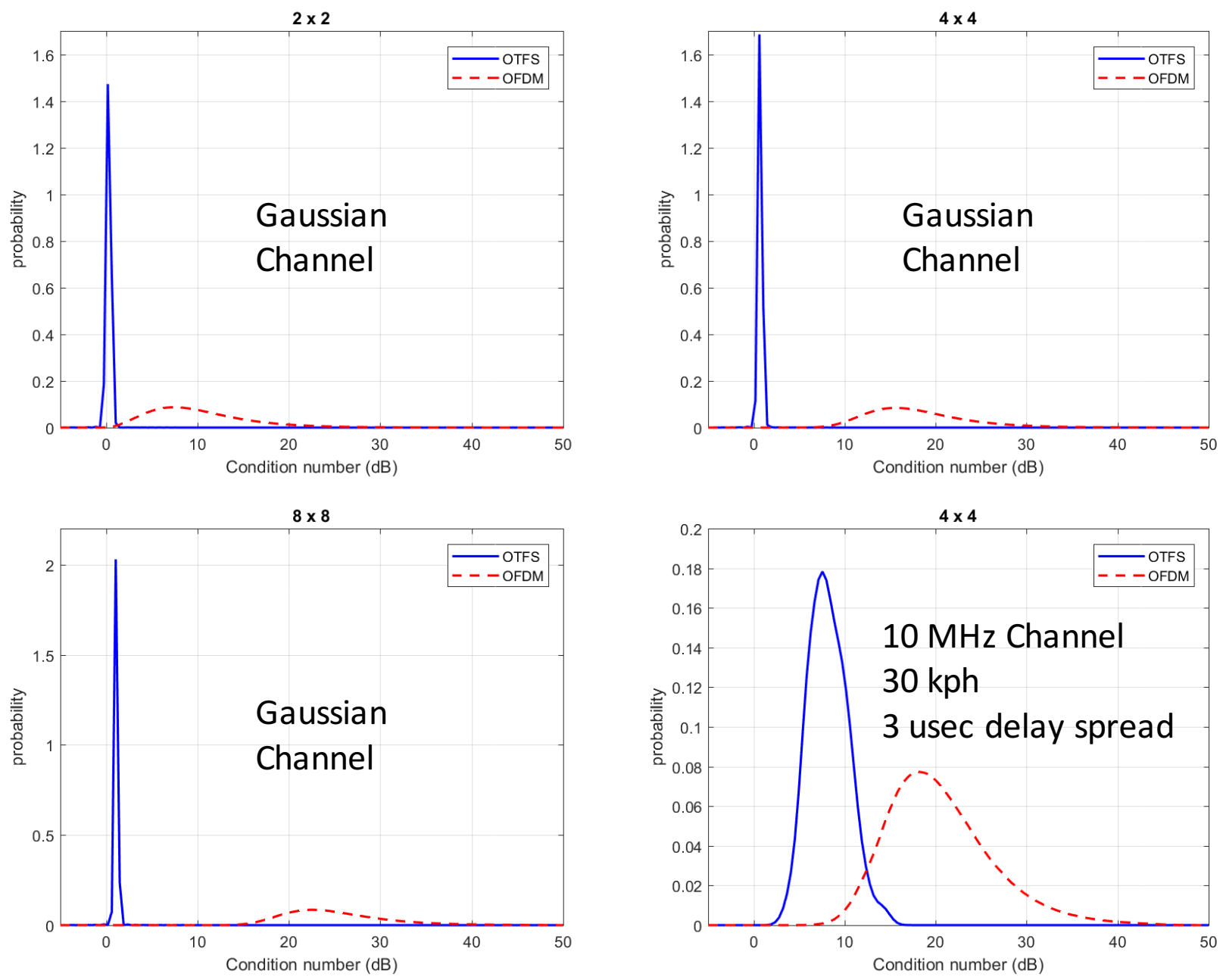

*Figure 15. Comparison of OTFS and OFDM condition numbers*

A simulation comparing the delay-Doppler condition numbers with the time-frequency condition numbers was carried out. The simulation considers two types of channels over a region of the time-frequency grid spanning a 20 MHz bandwidth and 10 sec duration. The first channel obeys a Gaussian model where every local matrix $\boldsymbol{U}_{m,n}$ is chosen independently, at random, for various MIMO orders. This channel model represents the unrealistic case of infinite delay and Doppler spreads. The second channel model is a realistic 4x4 MIMO channel with maximum Doppler spread corresponding to 30 km/h and maximum delay spread of 3 microseconds. Figure 15, shows the condition numbers of the time-frequency auto-correlation matrices $\boldsymbol{R}_{m,n}$, computed at each OFDM time-frequency grid point and plotted as a dashed histogram. In addition, the delay-Doppler condition numbers are computed for strips of 1ms duration and plotted as a solid histogram. The histograms clearly demonstrate that the average delay-Doppler condition number is significantly lower (better) than the average time-frequency condition number implying superior spatial multiplexing of OTFS over multicarrier modulation. Moreover, the variability of the delay-Doppler condition number is considerably smaller, implying increased consistency of performance. Simulation results comparing OTFS and OFDM with MIMO equalization are shown in Section 4.2.1.

## 3.2. DELAY-DOPPLER PRECODING

In the downlink direction, the received stream of each user is superimposed with interference induced by the other user's streams. Since, the antenna aperture of the user is of limited angular separation capabilities the interference from the undesired streams must be rejected at the base-station by means of pre-equalization (aka precoding). In the multicarrier setting the base-station is multiplexing a vector of $L_u$ QAM symbols over a region of the time frequency grid where each coordinate of the vector is reserved to a different user. Under these assumptions, the downlink channel decouples into a parallel system of simple MIMO channels over the time-frequency grid such that to every grid point $(m\Delta t, n\Delta f)$ there corresponds a local channel equation of the form:

$$Y_{m,n} = \boldsymbol{D}_{m,n} \cdot X_{m,n} + W_{m,n}$$

where $X_{m,n}$ is the transmitted vector of QAM symbols and $\boldsymbol{D}_{m,n}$ is the $L_u \times L_b$ matrix accounting for the local coupling between the $L_b$ antenna elements in the base-station and the $L_u$ user antennas. Each user receives its own signal corrupted by an interference induced by the signals directed to the other users. The standard method of rejecting the interference is called channel inversion or zero-forcing precoding (ZFP for short, [9], [10]). In this method, the base-station inverts the channel matrix $\boldsymbol{D}_{m,n}$ and transmits instead the pre-coded vector:

$$Z_{m,n} = \sqrt{\frac{NM}{\sum_{n,m} \left\| \boldsymbol{D}_{m,n}^{-1} \cdot X_{m,n} \right\|^2}} \boldsymbol{D}_{m,n}^{-1} \cdot X_{m,n}$$

The normalization constant ensures that the total transmission energy is normalized to $NM$. As a result, each user receives his pre-equalized QAM symbol corrupted by white noise with received SNR equal to:

$$SNR = \frac{NM}{N_0 \sum_{n,m} \left\| \boldsymbol{D}_{m,n}^{-1} \cdot X_{m,n} \right\|^2}$$

where $N_0$ is the variance of the noise term. In the presence of time and frequency selectivity, a considerable portion of the channel matrices $\boldsymbol{D}_{m,n}$ might be in in fade and as a result the corresponding normalization term $\left\| \boldsymbol{D}_{m,n}^{-1} \cdot X_{m,n} \right\|$ increases considerably. This in turns leads to SNR degradation. This phenomenon renders zero forcing precoding strictly sub-optimal. We note that variants of the ZFP using a regularized inverse of the channel can lead to small improvements to the received SNR [11]. To keep the exposition simple, we will restrict our attention to non-regularized ZFP for the remainder of the document.

A way to gain major improvements of the received SNR is by introducing a (lattice) perturbation to the transmitted QAM vector before it goes into the precoding filter. When the perturbation vector is properly chosen, the local normalization terms $\left\| \boldsymbol{D}_{m,n}^{-1} \cdot X_{m,n} \right\|$ can be reduced significantly. However, finding the optimal perturbation vector has exponential complexity, therefore in practice Tomlinson-Harashima Precoding (THP) is used to find a "good" perturbation vector [11], [12], [13]. We do not discuss the formal definition of the THP here, we only mention that the received SNR of ZF THP precoding critically depends on the condition number of the auto-correlation matrix:

$$\boldsymbol{R}_{m,n} = \boldsymbol{D}_{m,n} \boldsymbol{D}_{m,n}^{*}$$

When the condition number is high, the received SNR is low, and vice versa. In certain cases, the received SNR can be further improved using lattice reduction techniques, however, this introduces a significant complexity toll due to the need to recalculate the reduced lattice basis for each time/frequency coherence interval.

Just like for uplink reception, the performance of the ZF THP precoding can be significantly improved by multiplexing the vector of QAM symbols in the delay-Doppler representation where the channel-symbol coupling is invariant, separable and orthogonal, given by convolution with the delay-Doppler impulse response of the downlink channel. For the sake of explanation, we assume few simplifying approximations. We assume that the downlink channel decouples into a parallel system of identical MIMO channels over the points of the delay-Doppler grid such that to each grid point $(n\Delta\tau, m\Delta\upsilon)$ there corresponds a local channel equation of the form:

$$y_{n,m} = \boldsymbol{d} \cdot x_{n,m} + w_{n,m}$$

where $x_{m,n}$ is the vector of QAM symbols and $\boldsymbol{d}$ is an $L_u \times L_b$ matrix representing the global coupling between the $L_b$ base-station antennas and the $L_u$ user antennas. We further assume that the autocorrelation matrix $\boldsymbol{r} = \boldsymbol{d}\boldsymbol{d}^{*}$ is equal to the arithmetic average of all the local time-frequency auto-correlation matrices $\boldsymbol{R}_{m,n}$:

$$\boldsymbol{r} = \frac{1}{NM} \sum_{n,m} \boldsymbol{R}_{m,n}$$

Since the local channel matrices, $\boldsymbol{D}_{m,n}$, are roughly independent from one another, at least when considered over distinct coherence intervals, the average matrix $\boldsymbol{r}$ has a much lower condition number than a typical $\boldsymbol{R}_{m,n}$. Consequently, the ZF THP precoding achieves higher SNR which in fact can be shown to be very close to the capacity of the downlink channel. In addition, one can employ lattice reduction techniques to improve the SNR even further without compromising complexity since the reduced basis should be computed only once per frame of $NM$ QAM symbols, due to invariance of the delay-Doppler CSC. Performance comparison between time-frequency and delay-Doppler ZF THP precoding is given in Section 4.2.2.

# 4. OTFS PERFORMANCE ADVANTAGES OVER OFDM

## 4.1. KEY 5G USE CASES

In this chapter, we explore the performance advantages of an OTFS modulation scheme based on delay-Doppler multiplexing over an OFDM multicarrier modulation scheme based on time frequency multiplexing. We focus on demonstrating the performance gains that are intrinsic to the modulation structure and do not rely on specific implementation. We consider five central use cases of the emerging 5G technological premise which include:

- Enhanced Mobile Broadband (eMBB). This use case revolves around multi-user MIMO communication incorporating large numbers of antennas at the base-station as an enabler for serving large number users with maximum spectral re-use.
- High mobility communication. This use case revolves around the need to establish a reliable and consistent communication link between and to mobile recipients such as in the case of vehicle to vehicle communication (V2V or V2X) and in the case of high speed trains.
- Internet of Things (IoT). This use case revolves around the need to establish a communication link between a base-station and a very large number of small devices that operate under strict power constraints.
- Co-existence with Ultra Reliable Low Latency Communication packets (URLLC). This use case revolves around the need to support a transmission mode for high priority, low latency communication packets which are transmitted in an overlaid fashion over regular data packets, thus introducing significant narrowband interference.
- mm-Wave communication. This use case revolves around communication in the mm wavelength regime driven by the high demand for new available spectrum. Realizing a reliable communication link in these bands is challenging due to poor propagation properties of electro-magnetic waves and the high level of phase noise in these frequencies.

For each use case we characterize the underlying objective (or opportunity) and specify the main technical problem that needs to be resolved to realize the objective. We conclude with a brief theoretical explanation of the performance gain of an OTFS based solution over an OFDM based solution and back it up with simulation results.

## 4.2. ENHANCED MOBILE BROADBAND

A key premise of 5G Enhanced mobile broadband is the ability to dramatically increase capacity through incorporation of multiple antennas at the base station. This enables reuse of the available spectrum among a large number of users, a paradigm referred to as multi-user MIMO. In this section, we give simulation results comparing performance of OFDM (using LTE numerology and 3GPP evaluation assumptions) with OTFS. In the uplink direction, we focus on OTFS performance-complexity tradeoff gain of the base-station receiver and in the downlink direction, we focus on OTFS precoding gain. Unless stated otherwise, we assume the following simulation parameters:

- System Bandwidth: 10 MHz
- Channel Model: 3GPP TDL-C, 300 ns delay spread
- TTI Duration: 1 msec
- Channel Estimation: Ideal (as per 3GPP evaluation assumptions for 5G)
- FEC: LTE Turbo code
- Receiver: OTFS Turbo, OFDM-ML

### 4.2.1. Equalization Results

In Figure 16 and Figure 17, we compare OTFS and OFDM spectral efficiency. At each SNR point, the maximum modulation and coding scheme (MCS) that achieves the 3GPP operating BLER of 10% is selected. Figure 16 shows the spectral efficiency comparison for large packets (50 PRBs) and MIMO orders ranging from one to four (i.e., SISO, 2x2 and 4x4). The results for OFDM are obtained using maximum likelihood detection. A maximum likelihood receiver, while optimal for OFDM, has exponentially increasing complexity with the MIMO order hence, for higher-order MIMO, receivers are often implemented with reduced complexity algorithms exhibiting loss in performance. As the lower limit on OFDM performance we show the results with a simpler MMSE receiver (typically used to compare performance in 3GPP for OFDM systems).

The gap between OTFS and OFDM is clearly seen and is particularly pronounced for higher order MIMO. For example, for 4x4 MIMO, at around 19 dB SNR, the performance gap ranges from 36% to 53%, depending on the type of OFDM receiver. Since the Doppler at 30 km/h is relatively low, and the packet size large, the gain is not from the additional diversity obtained due to the spreading effect of OTFS. Rather, it is due to the condition number argument presented in Section 3.1. Further gain is seen in Figure 17. This is due to the smaller packet size. In OTFS, performance is invariant to packet size since all symbols experience the full diversity of the channel. In contrast, a small OFDM packet is more likely to be ‘stuck’ in a time and/or frequency-selective fading region, relying on the FEC code to recover.

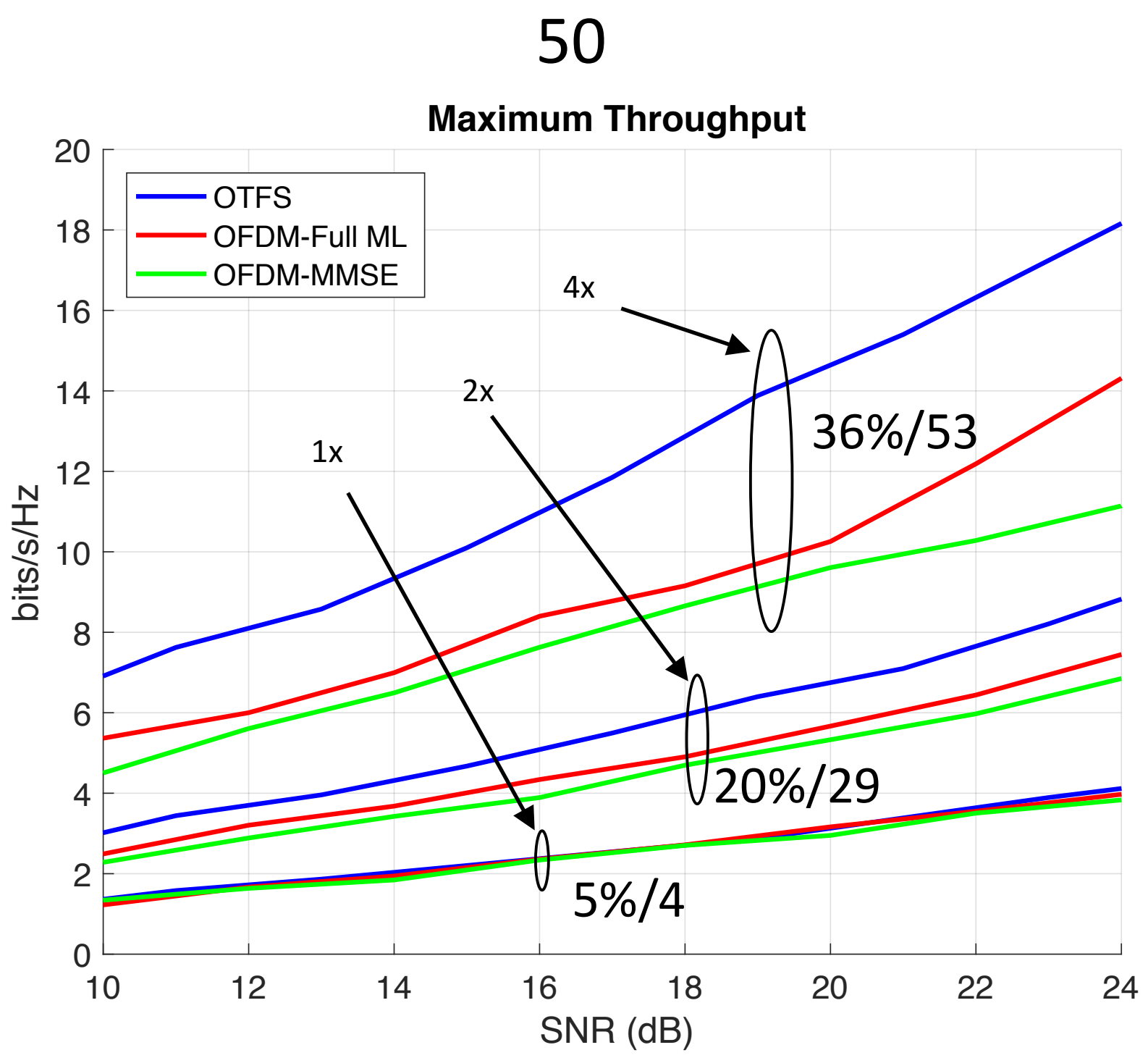

*Figure 16. Large packet throughput: 30 km/h*

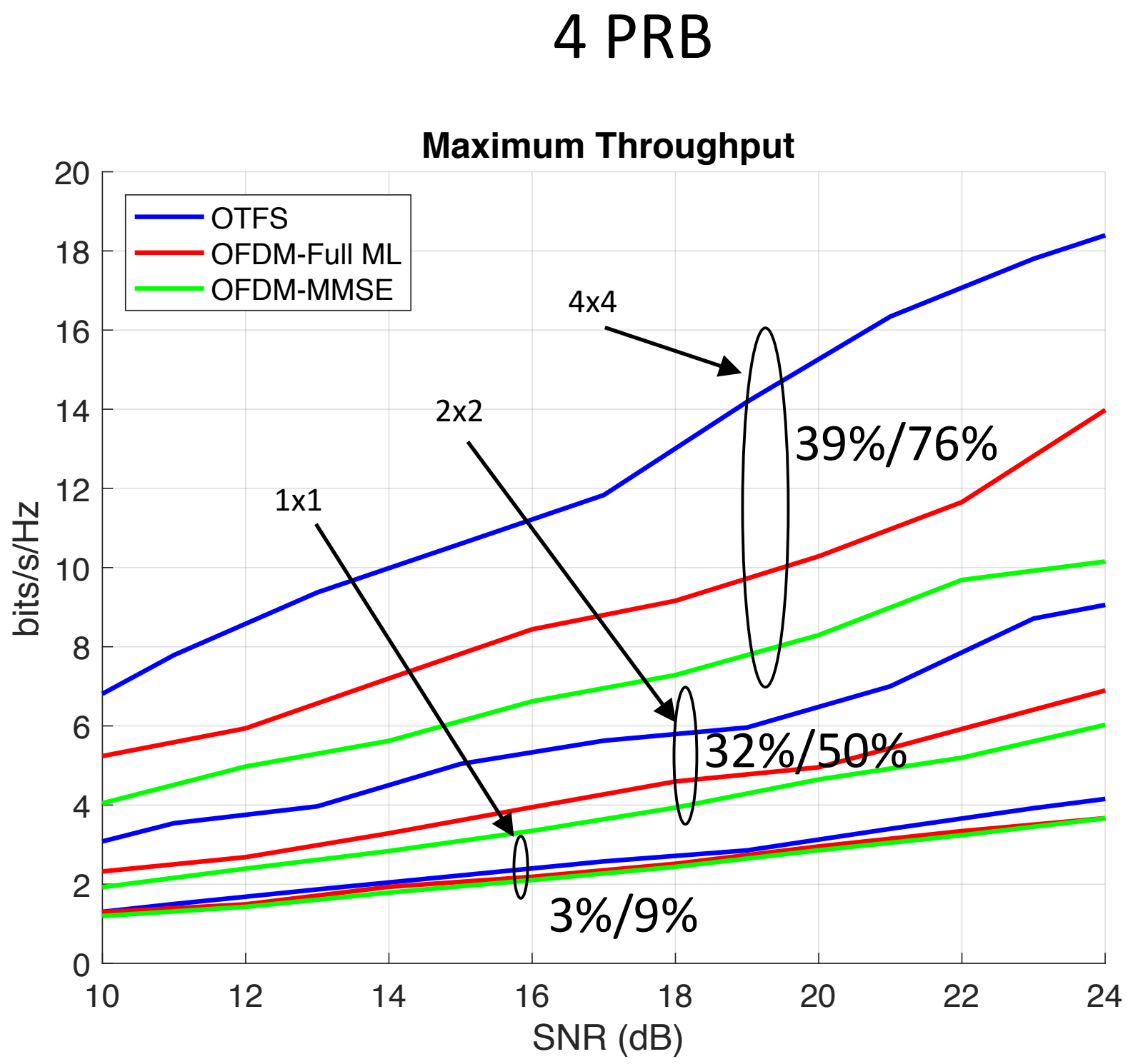

*Figure 17. Small packet throughput: 30 km/h*

### 4.2.2. Precoding Results

To evaluate the performance gain of delay-Doppler over time-frequency zero-forcing Tomlinson Harashima precoding (ZF THP), we run a simple simulation of a wireless cell of 1 km radius containing several thousand users randomly located, each equipped with a single antenna, and a base station consisting of a linear array of 8 antenna elements. We distribute around each user a ring of static reflectors corresponding to a delay spread of 2 microseconds and no Doppler. At every iteration, we choose at random a subset of 8 users and compute both time-frequency and delay-Doppler ZF THP received SNR for each user. The experiment is repeated several thousand times with different configurations of users and reflectors. The cumulative distribution function of the received SNR values is shown in Figure 18. The simulation uses the following parameters:

- RF Frequency: 4 GHz
- RF Bandwidth: 10 MHz
- Subcarrier spacing: 15 kHz
- Antenna spacing: 20 cm
- Packet duration: 1 msec

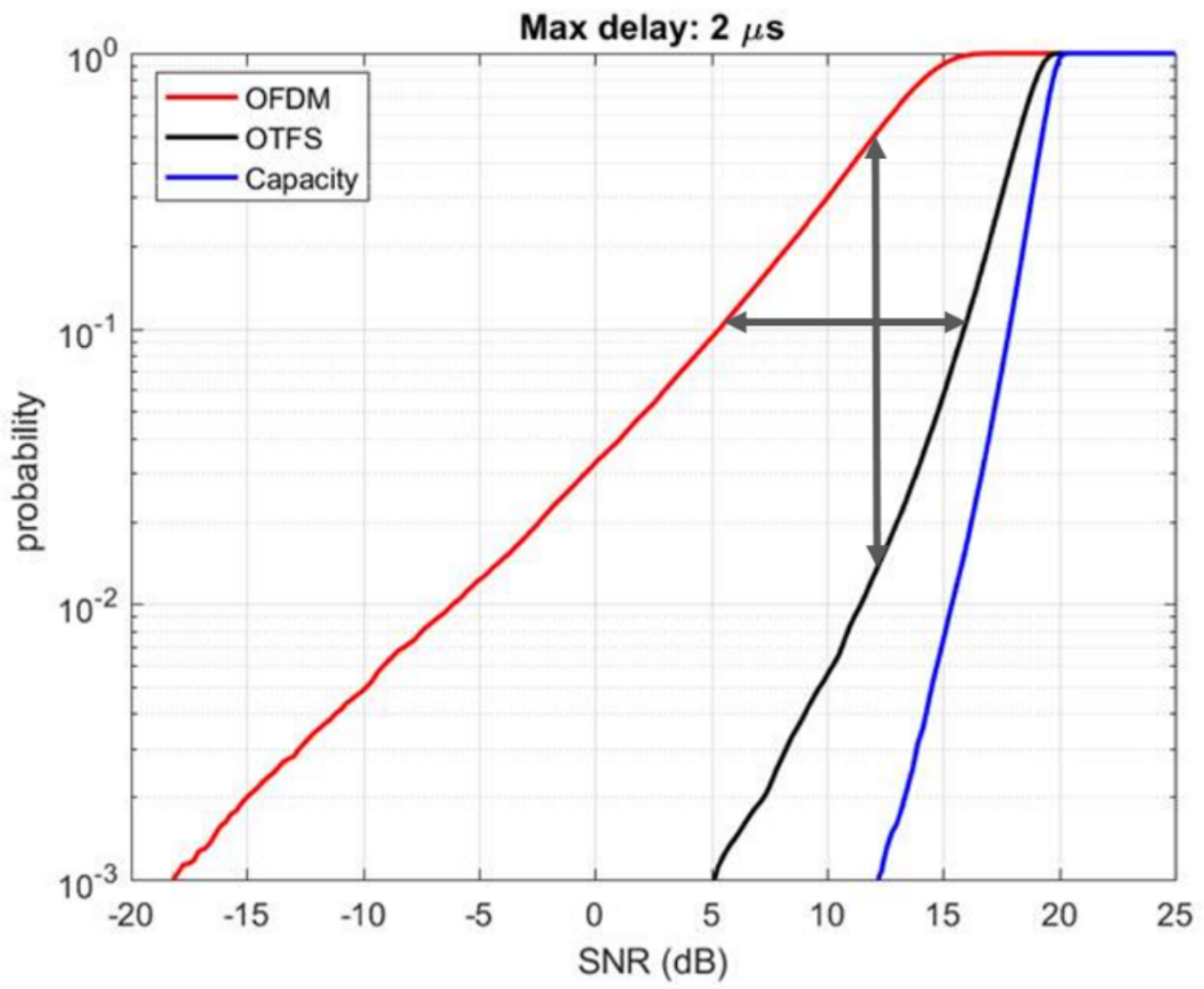

*Figure 18. Precoding Comparison*

The vertical line in Figure 18 shows that around 99% of OTFS users compared with around 50% of OFDM users enjoy an SNR greater than 12dB. In other words, 99% of OTFS users experience the performance of the top 50% of the OFDM users. The horizontal line designates that 90% of OTFS users has more than 10dB SNR gain compared with 90% of the OFDM users.

## 4.3. COMMUNICATION UNDER HIGH MOBILITY CONDITIONS

Communication under mobility conditions includes use cases of extreme mobility where either the transmitter or the receiver are moving (in contrast to the fixed situation where both the transmitter and the receiver are static and the only moving objects are the reflectors). Typical scenarios are communication between a vehicle and another vehicle (V2V), communication between a vehicle and a static base-station or infrastructure (V2I), communication between a base-station and a drone, communication between a base-station and a fast-moving train, and many more. The high mobility communication channel is characterized by wide Doppler spread.

The principal objective of operation under high mobility conditions is to maintain a reliable and consistent communication link supporting predictable performance to many users for various packet sizes. There are two main technical challenges. The first is concerned with the presence of intercarrier interference (ICI) due to Doppler which results in SNR degradation. The second challenge is related to the short coherence time scale of the channel which results in unpredictable fluctuations in the temporal power profile and phase of the received signal, rendering adaptation of the allocated subcarriers and MCS unrealistic.

Figure 19 shows a comparison of OTFS and OFDM with a typical set of 3GPP evaluation assumptions, namely 300 ns RMS delay spread, 120 km/h and 2x2 MIMO for 16 QAM and 64 QAM with large packets. The graphs show that the performance gap at 10% coded BLER is significant, ranging from 2.4 dB to 4 dB, due to the additional Doppler/time diversity gain of OTFS over OFDM.

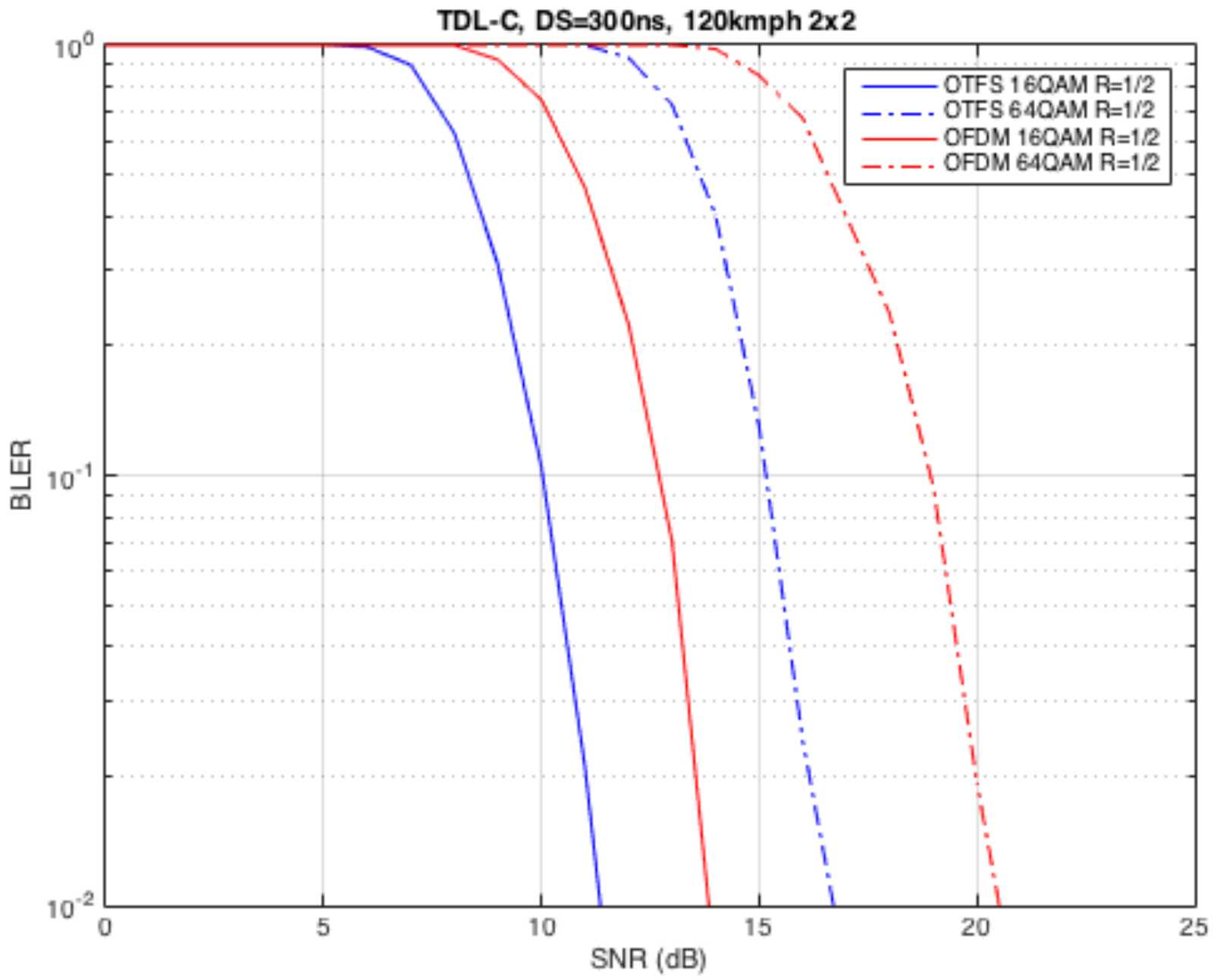

*Figure 19. Moderate speed use case*

Figure 20 shows the performance in a high-speed train (HST) use case at 500 km/h. At such high speeds, the Doppler spread is a significant percentage of the subcarrier spacing (SCS) and induces non-negligible intercarrier interference (ICI). A common method for improving performance is to increase the SCS. This improves performance for both OFDM and OTFS however, as can be seen in the figure, OTFS with 15 kHz SCS (i.e., no change) outperforms OFDM with 60 kHz SCS by about 2.6 dB. The ability to operate at a smaller SCS has further implications. Increasing the SCS reduces the OFDM symbol size. However, the length of the cyclic prefix (CP) depends only on the delay spread of the channel, hence if the symbol size decreases by a factor of four and the delay spread does not change, the resulting CP overhead is increased by the same factor thus further reducing the effective throughput (which is not represented in the BLER plots). In other words, OTFS has the dual benefit of improved BLER performance (translating to higher MCS) and lower CP overhead, as compared with OFDM in this scenario.

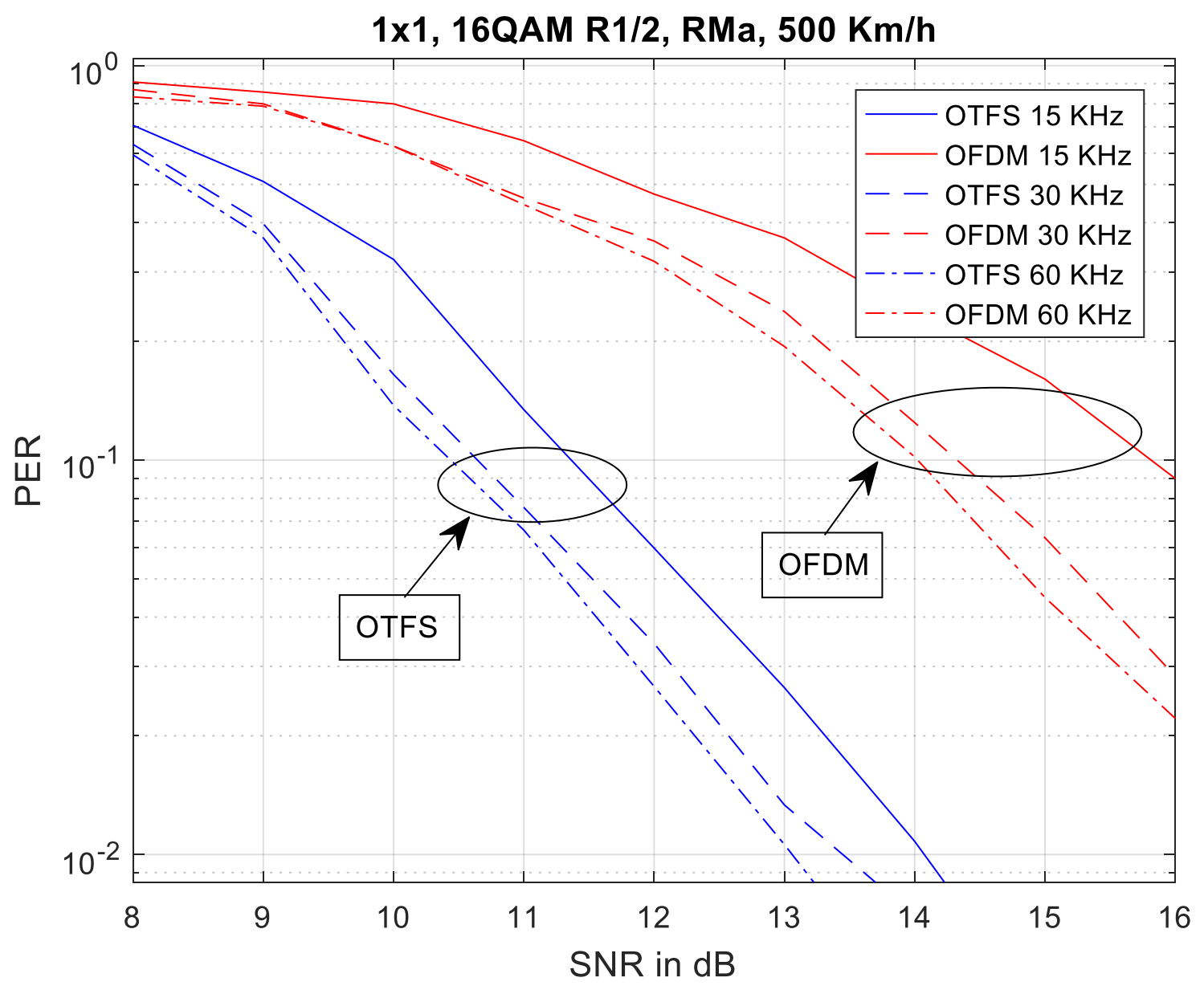

*Figure 20. High speed use case with variable subcarrier spacing*

Figure 21 further expands on the HST use case, highlighting the effect of ICI cancelation. In this example, we simulate an ideal ICI cancellation by completely removing the large ICI impact. The graph shows that the performance gap of OTFS is significant, and results from capturing the full diversity of the channel and not just due to ICI resilience. As can be seen in this SISO case, a 64 QAM OFDM signal without ICI cancelation is completely degraded and unable to even achieve the required 10% BLER. In contrast, OTFS still achieves this at around 22 dB SNR. We also see that with complete ICI cancelation, the SNR gap for 64 QAM is about 4 dB.

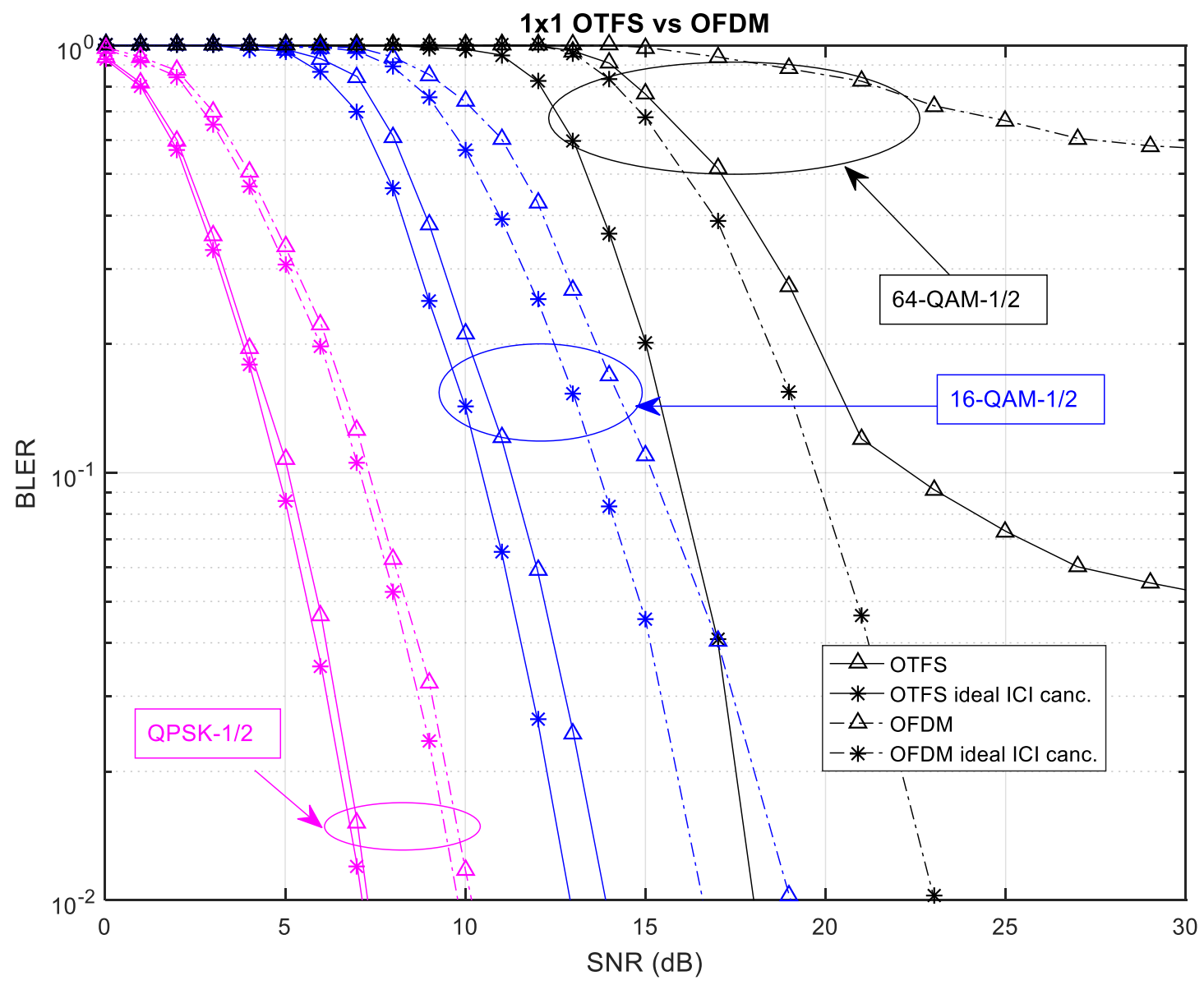

*Figure 21. High speed use case (500 km/h) with ICI cancelation*

The devastating effect of lack of adaptation on the performance of multicarrier modulations based on time-frequency allocation is especially apparent for small packet size. Such packets, when designated to an arbitrary region of the time-frequency grid without taking into consideration the channel condition at this region, have a non-zero probability of being affected by a deep fade. In this event, the information in the packet is lost without any regard to the specific error correction and receiver structure being used.

In lack of adaptation, the standard approach to mitigate the fading phenomenon in the multicarrier setting is through application of interleaving and coding. In this approach the information bits are interleaved over non-continuous regions of the time-frequency grid spanning multitude coherence bandwidths and time intervals and use of the error-correction code to extract the diversity of the channel. This works well for big packets which can span many coherence intervals and use long codewords to overcome the fading event. However, for small packets using shorter codewords and spanning small number of coherence intervals, the effectiveness of this approach degrades considerably.

In contrast to time-frequency multiplexing which allocate a region of the time-frequency grid of the size of the packet, OTFS multiplexes the packets over the delay-Doppler domain. In this multiplexing method, every QAM symbol is spread over the full time-frequency grid thus is affected by all the diversity modes of the channel resulting in a diversity gain that is independent of the packet size. In terms of system performance, this translates to increased throughput consistency which accentuates with the incorporation of higher layer TCP protocol. Figure 22 compares OTFS with OFDM based LTE transmission for small packets size, consisting of 4 PRBs, in mobility channels corresponding to a speed of 30 km/h.

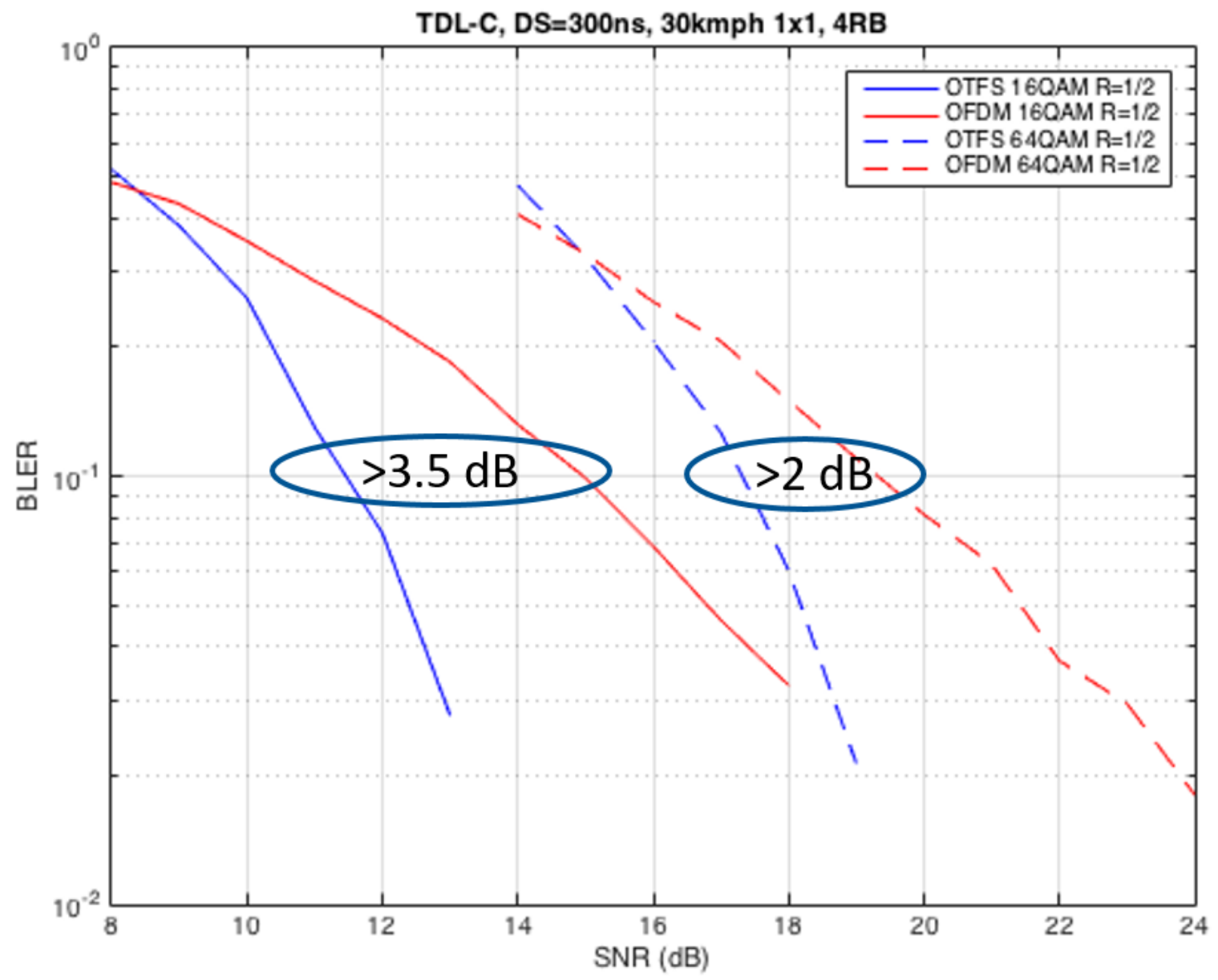

*Figure 22. Small packet use case*

## 4.4. COMMUNICATION UNDER NARROWBAND INTERFERENCE

A key use case of 5G network revolves around ultra-reliable, low-latency communications, which includes applications such as industrial internet, smart grids, infrastructure protection, remote surgery and intelligent transportation systems. Meeting this use case requires the network to support the option for an abrupt transmission of low latency, small communication packets used for high priority signaling. The transmission protocol of the URLLC packets is to overlay them over regular data packets by puncturing small segments and installing in-place the URLLC content. There are two methods to achieve this: one is when the receiver is notified ahead of time about the location and size of the URLLC intruding packet (Indicated URLLC) and the other is when the receiver is not informed about the presence the URLLC packet (Non-indicated URLLC).

The presence of a parasitic URLLC packet introduces a narrowband additive interference to the hosting data packet that can significantly affect the performance of the receiver. However, the devastating effect of each of the two transmission modes on the overall performance is distinctively different in OFDM. In the Indicated transmission mode, the interference of the URLLC packet is known to the receiver (at least in terms of its position) and hence the main data packet can be decoded by deliberately ignoring the information located at the designated affected area of interference. At least for large data packets, this loss of received signal can be compensated by the FEC so that data recovery is not compromised. This resembles the way data is recovered in the presence of channel fading, where the receiver uses the channel state information to locate and ignore the signal in the faded regions and compensates for the loss of the received signal using the FEC.

The Non-indicated mode is reminiscent of operation under unknown additive narrowband interference and thus presents a more serious problem. In this situation, due to the lack of knowledge about the location of the interference, the receiver is not able to disregard the parasitic bits which cause a systematic confusion in the FEC decoding cycle. This results in a significant reduction in performance with little regard to the relative size of the packet and the rate of the code.

In multicarrier modulation, the URLLC bits directly interfere with the data bits, thus resulting in a total confusion of the FEC decoding cycle. In OTFS, however, the data information bits are residing on the dual delay-Doppler grid and, prior to FEC decoding, the URLLC interference bits are spread over the whole delay-Doppler grid by the symplectic Fourier transform. The resulting effect is merely a small SNR degradation.

Figure 23 depicts simulated PER comparison between OTFS and OFDM based LTE transmission for large data packets in the presence of URLLC interference. The simulation covers both the Indicated and Non-indicated modes. OTFS, as a spreading technique, enjoys intrinsic resilience to narrowband interference for both Indicated and Non-indicated URLLC (in the same spirit as CDMA), while in contrast OFDM is very sensitive to this type of additive impairment. In the Non-Indicated mode, OFDM breaks down completely in the presence of URLLC packets while OTFS suffers between 0.5 dB and 2 dB of degradation. In the Indicated transmission mode (adopted by 3GPP) OFDM exhibits almost 2 dB of degradation relative to OTFS and then suffers increasingly worse relative degradation in the presence of URLLC.

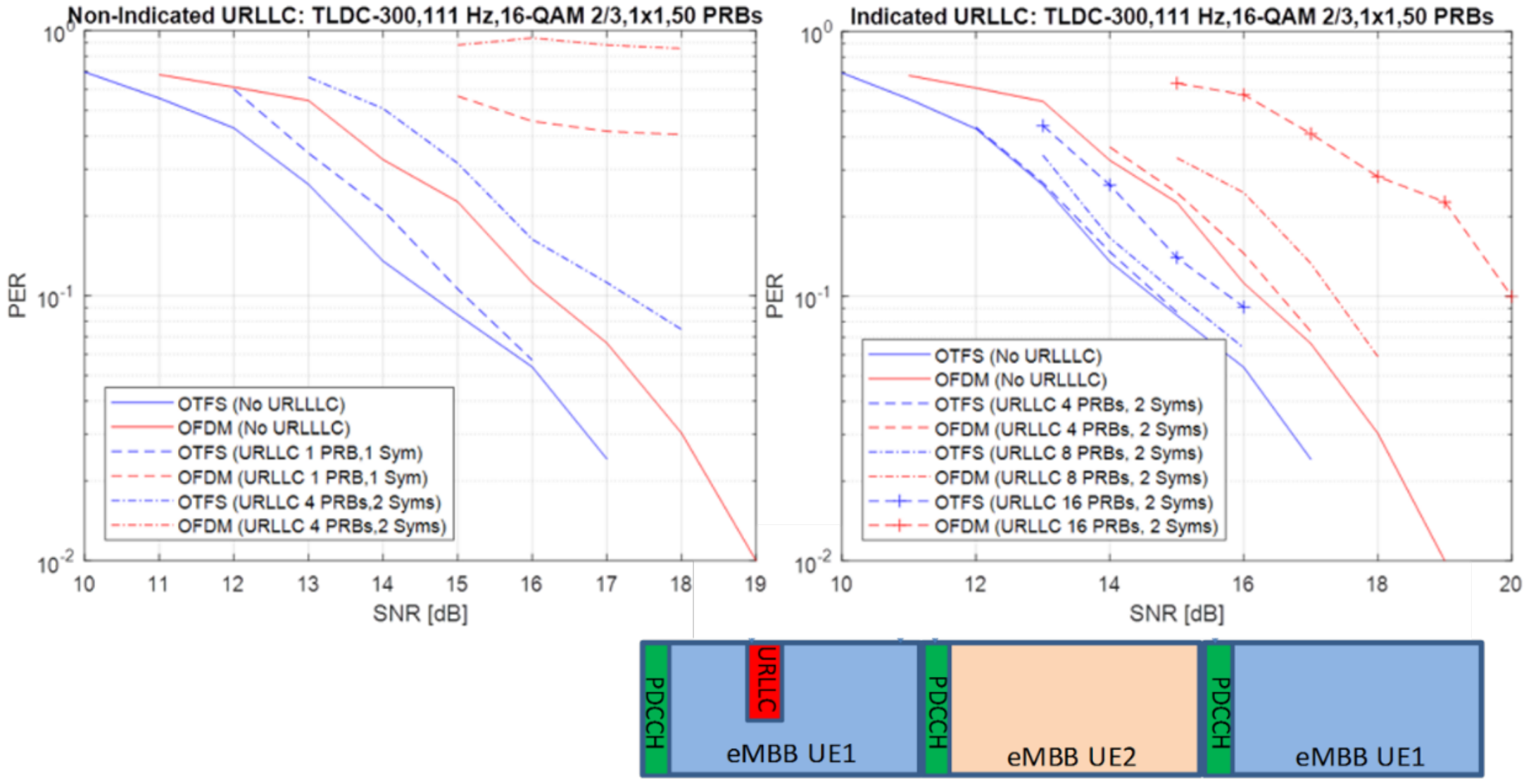

*Figure 23. URLLC Coexistence - Non-Indicated vs Indicated*

## 4.5. COMMUNICATION UNDER POWER CONSTRAINTS – INTERNET OF THINGS

To date, the wireless network has supported mainly voice calls and data services, all revolving around human recipients. Internet of Things (IoT) is a synonym for a major 5G use case that revolves around massive machine type communication (mMTC for short) between billions of devices that are expected to be connected to the wireless network. These devices generally transmit small packets and operate under strict transmit power constraint to extend their battery life. The power constraint imposes a formidable challenge in achieving in-building penetration and extended coverage.

The main technical challenge is concerned with the need to maximize the link budget and minimize the number of retransmissions (energy per bit of information) under transmit power constraint and latency requirement. To maximize the link budget under these constraints one should reduce the Peak to Average Power Ratio (PAPR) of the transmitted signal and maximally extend the duration of the transmission under the latency requirement. To minimize the number of retransmissions one should extract time and frequency diversity gain. To optimize performance the transmitted waveforms should simultaneously meet the following criteria:

- Minimum PAPR.
- Maximum diversity gain.
- Maximum transmission duration.

There are two basic approaches to meet these criteria in multicarrier modulation. The first approach, referred to as single carrier FDMA (SC-FDMA), multiplexes the data using a DFT spread along a fixed narrow bandwidth repeated over multiple of multicarrier symbols for the full duration allowed by the latency requirement, see the left part of Figure 24. This mode of transmission maximizes the link budget as it achieves low PAPR comparable to that of single carrier modulation and maximum transmit duration. However, due to its narrowband nature it is susceptible to frequency selective fading (low frequency diversity gain) thus suffering a high average number of retransmissions.

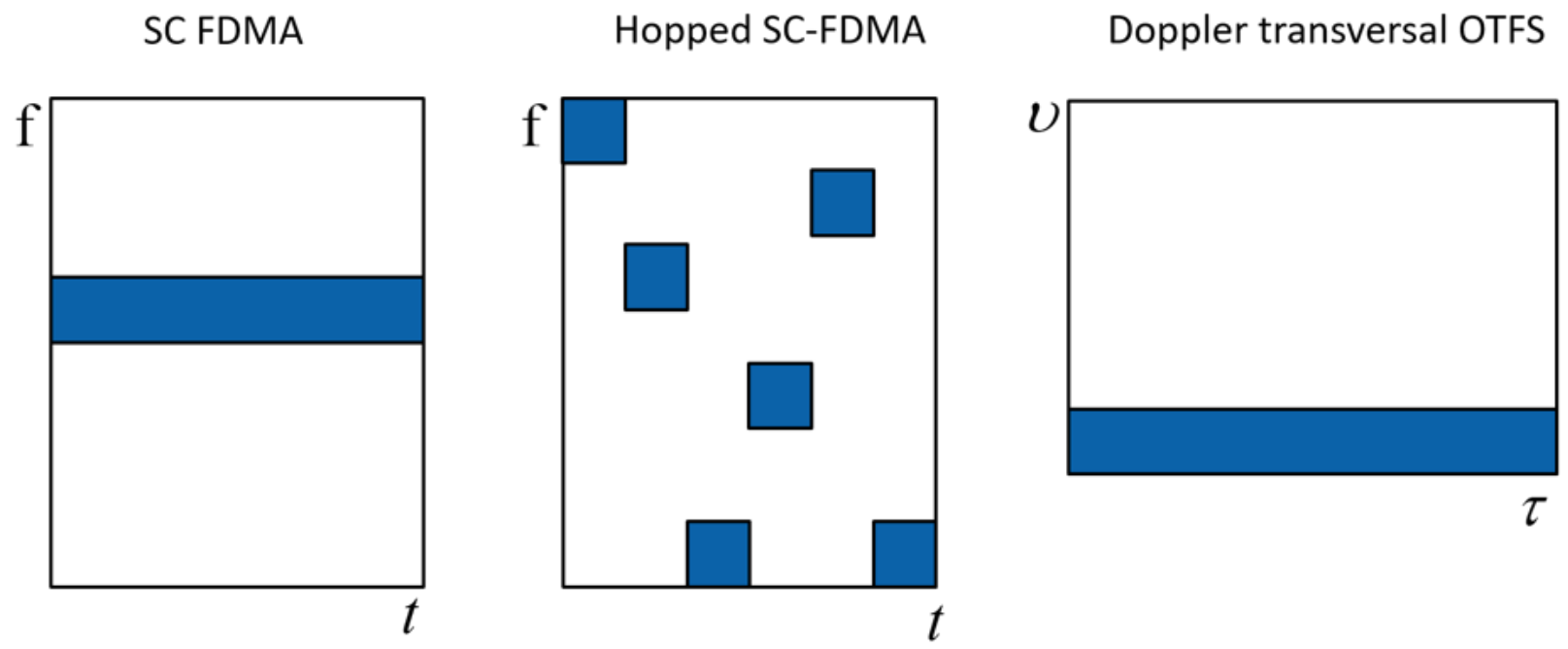

*Figure 24. From left to right: SC FDMA, Hopped SC FDMA and Doppler transversal OTFS packet allocations*

A more sophisticated variant is referred to as hopped SC-FDMA. In this variant, one still applies DFT spreading over narrowband portions of the spectrum, but instead of using a fixed band, one jumps between multiple bands to exploit multiple channel modes (see the middle part of Figure 24). This mode of transmission maximizes the link budget as it enjoys low PAPR comparable to single carrier and maximize transmission duration, while at the same time extracting additional diversity gain. However, there is a subtle phenomenon that renders this approach sub-optimal. To maintain low PAPR, the QAM order must be kept low - say QPSK. Under this constraint, the transmission rate can only be adjusted by changing the FEC rate[5]. Thus, the performance is governed by the restricted QPSK capacity (or restricted mutual information) instead of by the Gaussian capacity. Unlike Gaussian capacity, the restricted capacity depends on the modulation scheme. A fundamental result in information theory [14] shows that in the presence of time-frequency selectivity the restricted capacity of multicarrier modulations is *saturated*, becoming strictly sub-optimal. Due to capacity saturation, hopped SC-FDMA requires a higher transmission power to support a fixed transmission rate.

To summarize, keeping the QAM order fixed, there is a fundamental limitation to simultaneously maintain low PAPR and extract diversity gain with multicarrier modulations. This fundamental limitation can be overcome by multiplexing the QAM symbols in the delay-Doppler representation. A simple analysis of the Zak transform reveals that allocating the information QAM symbols along a single Doppler coordinate (referred to as Doppler transversal allocation and shown in the right part of Figure 24),

[5] In contrast to regular multicarrier data transmission modes which maintain low coding rate at the expense of increasing the QAM order

simultaneously achieves maximum link budget (as it enjoys low PAPR and maximum transmit duration) and extracts full time-frequency diversity while avoiding the restricted capacity saturation phenomena due to the convolutive delay-Doppler channel-symbol coupling. The use of Doppler transversal allocation renders OTFS the absolute optimal modulation for maximizing link budget and minimizing the number of retransmissions.

We conclude this section with simulation results. The first simulation compares the PER performance of OTFS using Doppler transversal allocation with SC-FDMA. Figure 25 depicts the PER performance gain of OTFS over SC-FDMA due to frequency diversity gain. Note that both modulations enjoy the same low PAPR comparable to that of single carrier and extend their transmission burst over several OFDM symbols to maximize link budget. The graph shows that to achieve PER of 1% at modulation order of QAM 64, OTFS requires almost 8dB less transmission power.

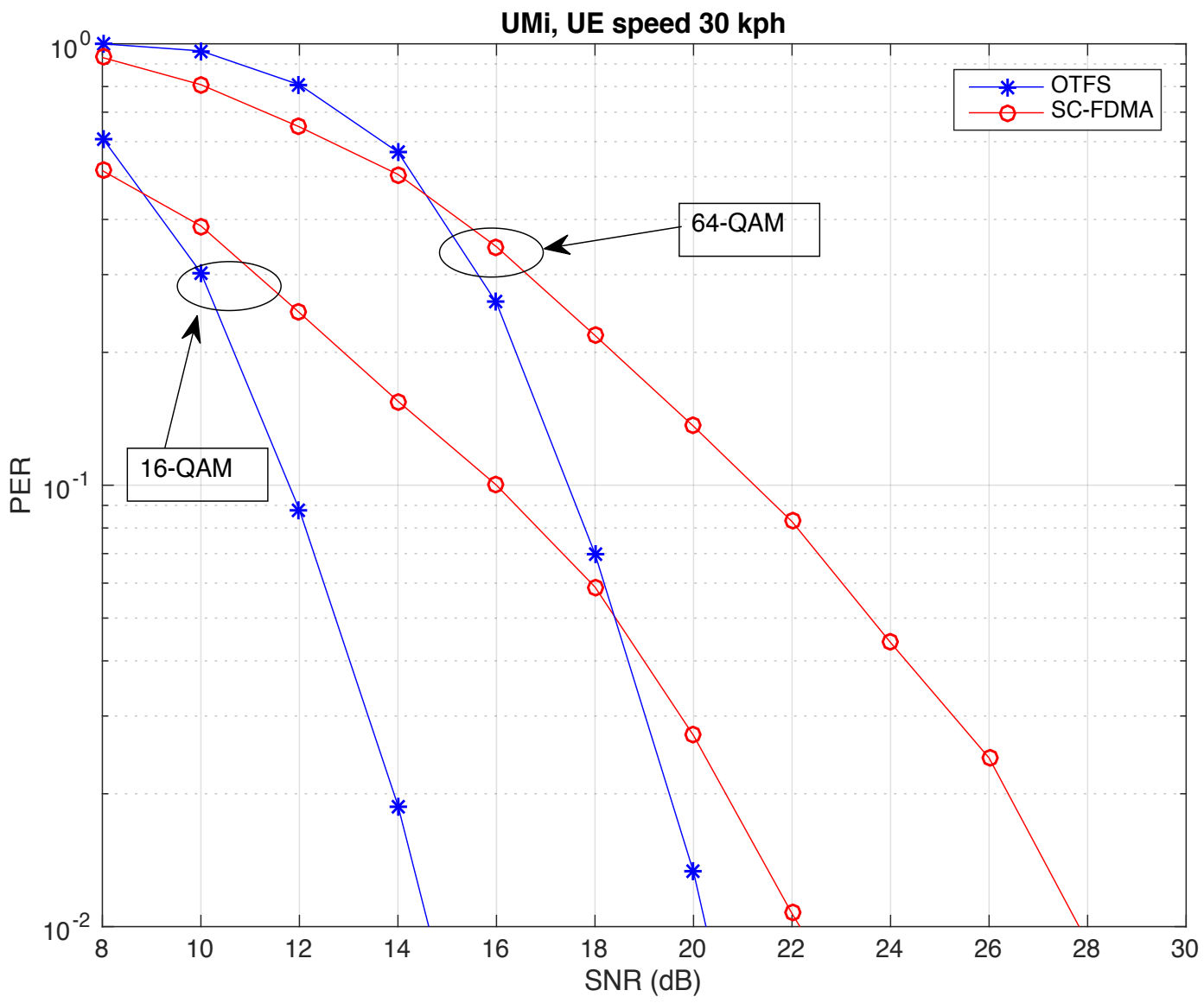

*Figure 25. Low PAPR Performance Gain*

The second simulation compares the performance of OTFS Doppler transversal allocation to hopped SC-FDMA for single antenna transmission of a single PRB sized packet consisting of 12x14=168 QPSK symbols at a high code rate of 0.9 through a frequency selective channel. Figure 26 depicts the PER performance gain of OTFS due to the restricted capacity saturation phenomenon. The graph shows that at packet error rate of 1%, OTFS requires 7dB less transmission power.

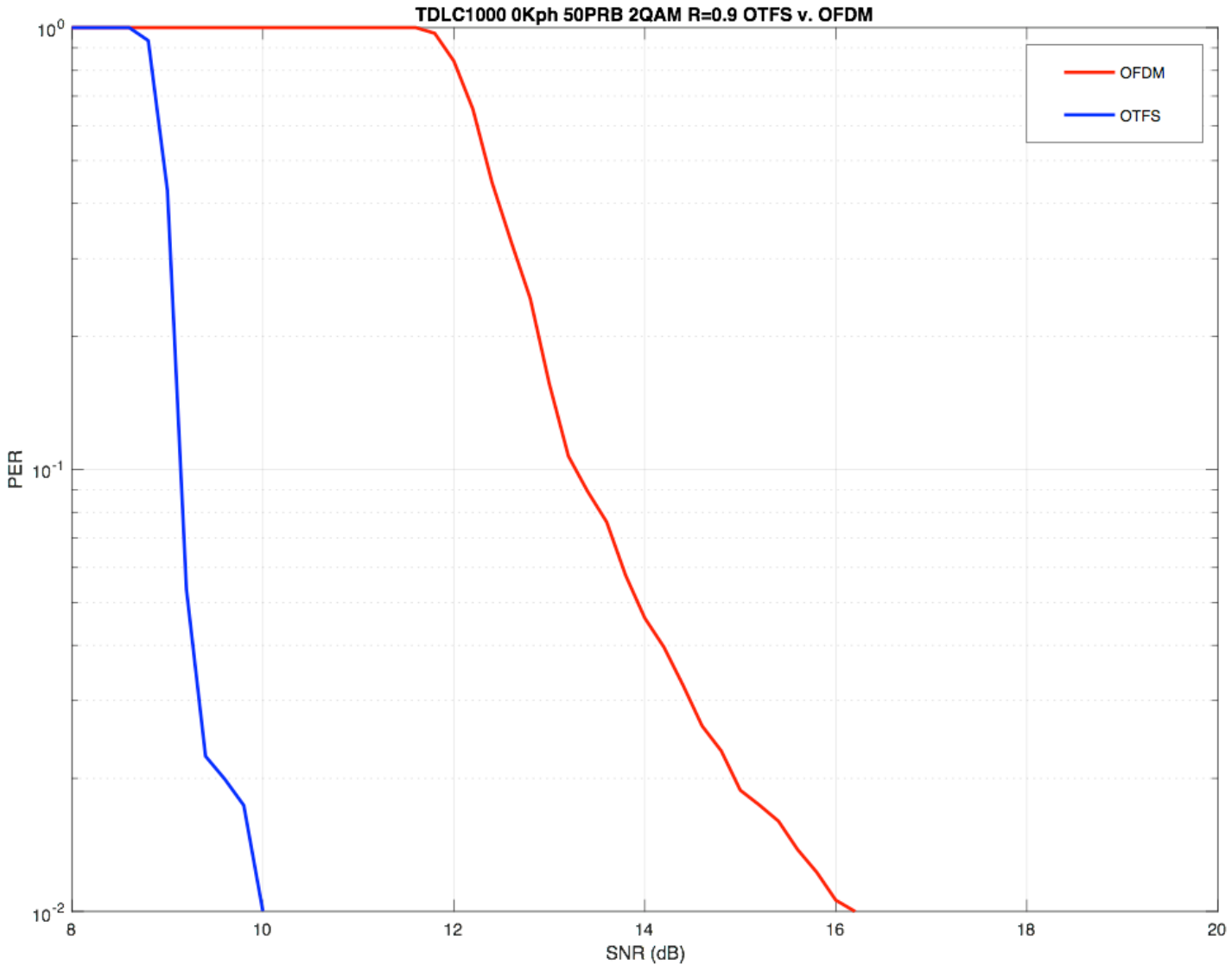

*Figure 26. Restricted Capacity Saturation*

## 4.6. MM-WAVE COMMUNICATION

The large spectrum availability in the millimeter wavelength regime opens an opportunity for significant upscaling of throughput. Communication at mm-wave frequencies is thus a major driver in the evolution of the emerging 5G network. Designing a scalable, cost effective communication system that operates at these high frequencies is a non-trivial task, however.

There are two basic technical challenges that need to be addressed. The first challenge is concerned with the power attenuation of electro-magnetic propagation in the mm-wave regime compared to the conventional cm wavelength (sub 6 GHz) commonly used in contemporary networks. A direct way around this issue is to maintain line of sight propagation conditions. This imposes, however, a severe restriction on the network architecture requiring the installation of many additional base-stations for network densification and thus leading to massive increases in capital expenditures. The second technical challenge is concerned with the RF oscillator phase noise that is significantly accentuated at high frequencies. The main issue associated with this effect is appearance of significant intercarrier interference (ICI) among adjacent tones which results in SNR degradation.

There are two approaches for mitigating ICI impairment in the multicarrier setting. One approach is to incorporate an interference cancellation mechanism at the receiver. The drawback of this approach is that it considerably complicates the receiver structure and, in addition, requires knowledge of the ICI coefficients, thus introducing an additional capacity overhead devoted for channel acquisition. The other approach is based on mitigation instead of cancellation. In this approach, the ICI effect is diminished by increasing the subcarrier spacing between adjacent tones. At high carrier frequencies, the expanding factor can be 10-20-fold compared to conventional LTE. The drawback of this approach is that increasing the subcarrier spacing leads to shortening the multicarrier symbol time by the same factor. Since the duration of the cyclic prefix (CP) depends solely on the channel delay spread, shortening the symbol time may lead to a proportional increase in CP overhead resulting with lower spectral utilization, as described in Section 4.3.

Inclusion of a cyclic prefix in the formation of the waveforms is used by multicarrier modulations to maintain orthogonality of the channel symbol coupling. In contrast, as explained in Section 2.5, delay-Doppler multiplexing does not *require* a cyclic prefix[6] to maintain the desired attributes of the CSC (stationarity, separability and orthogonality. In fact, one can expand the Doppler period to an extent which allows the acquisition of the phase noise as part of the channel state information and then the cancelation of its effect through equalization. This method avoids any sacrifice in capacity due to unaccounted interference or reduced spectral utilization due to CP overhead. Intuitively, this capability can be seen in the structure of the waveform shown in Figure 8.

Figure 27 depicts the results of a simplified simulation showing the gain in spectral efficiency of OTFS (with no CP) compared to OFDM under the presence of phase noise impairment characteristic to the mm-wavelength regime. As can be seen, OTFS spectral efficiency outperforms that of OFDM by 2-3 bits/sec/Hz, depending on the delay spread of the channel primarily due to the combined degrading effect of the CP overhead and ICI on the performance of OFDM.

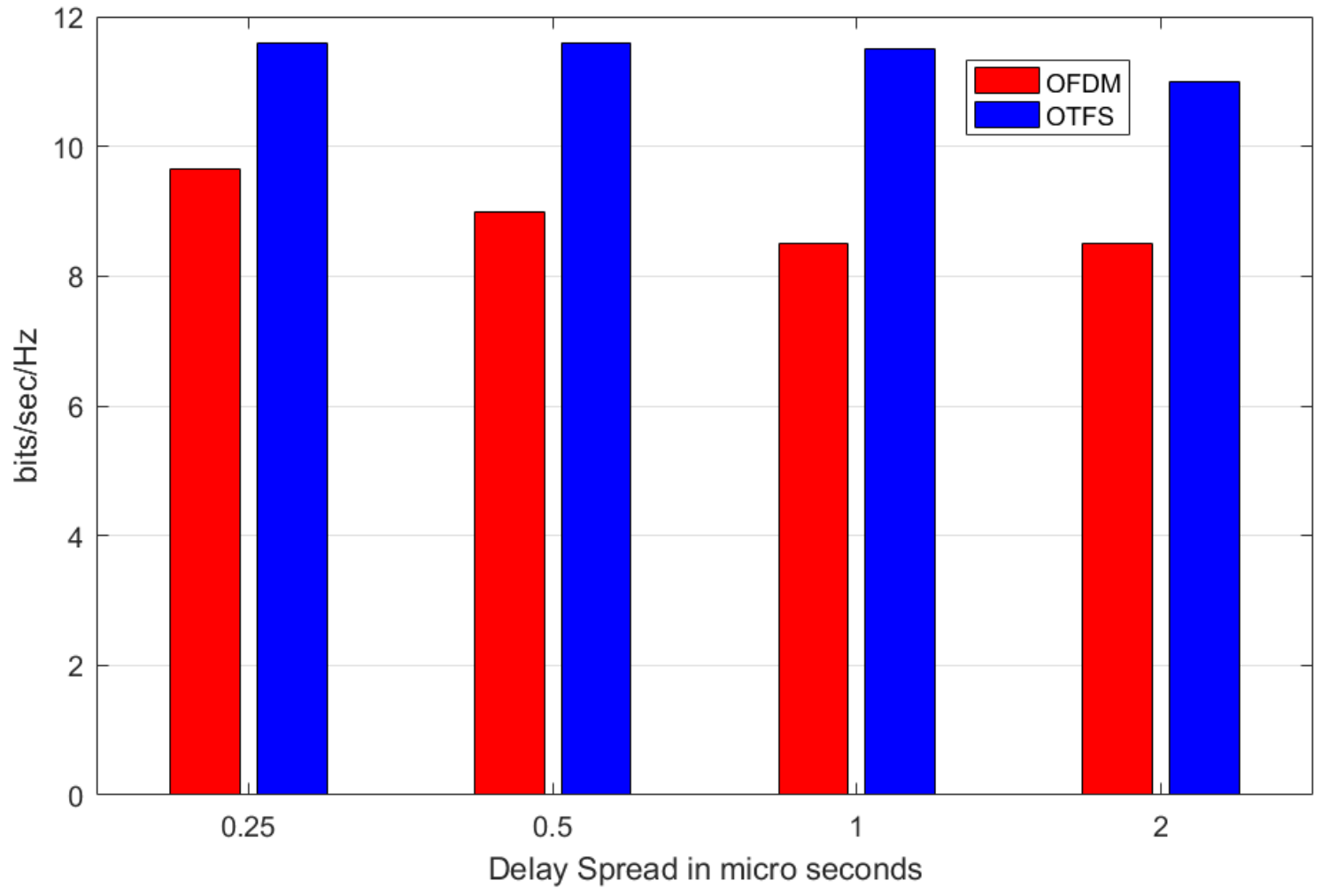

*Figure 27. mm-Wave Capacity Gain*

[6] Recall that we described a multicarrier version of OTFS in Section 2.7 to easily coexist with a multicarrier modulation solution, however this is not a necessary requirement for OTFS modulation

# 5. SUMMARY

OTFS is a novel family modulation scheme based on multiplexing the QAM information symbols over localized pulses in the delay-Doppler signal representation. The OTFS modulation schemes constitutes a far reaching generalization of traditional time and frequency modulation schemes such as TDMA and OFDM, which can be shown to be limiting cases of the OTFS family.

From a broader perspective, OTFS establishes a conceptual link between Radar and communication. The OTFS waveforms couple with the wireless channel in a way that directly captures the underlying physics, yielding a high-resolution delay-Doppler radar image of the constituent reflectors. Thus, the time-frequency selective channel is converted into an invariant, separable and orthogonal interaction, where all received QAM symbols experience the same localized impairment and all the delay-Doppler diversity branches are coherently combined.

The OTFS channel-symbol coupling allow linear scaling of capacity with the MIMO order while satisfying an optimal performance-complexity tradeoff both at the receiver end (using joint ML detection) and at the transmitter end (using Tomlinson Harashima precoding for MU-MIMO). OTFS enables significant spectral efficiency advantages in high order MIMO under general channel conditions over traditional modulation schemes including multicarrier modulations such as OFDM.

OTFS can be viewed as a special type of a time-frequency spreading technique, where each QAM symbol is carried by a two-dimensional basis function spread over the full time-frequency grid. When viewed as a time-frequency spreading technique, OTFS exhibits architectural compatibility with any type of multicarrier modulation scheme, including conventional OFDM. An OTFS packet can be flexibly designed to populate arbitrary regions of the time-frequency grid and be compatible with any convention for channel reference signaling. As a spread spectrum, OTFS enjoys resilience to narrowband interference and full diversity gain.

OTFS resilience to interference makes it ideal to supporting ultra-reliable low latency communication packets overlaid on regular data packets. OTFS diversity gain makes it ideal for communication under mobility conditions.

OTFS supports a small packet allocation method (called Doppler transversal allocation) that maximizes link budget and minimizes number of retransmissions under transmit power and latency constraints by achieving the PAPR of single carrier, extracting full time-frequency diversity and maximizing restricted capacity. OTFS, with Doppler transversal allocation, is superior to conventional multicarrier DFT spread techniques such as SC-FDMA and its more elaborate hopped variant for applications of IoT.

3GPP has identified a variety of eMBB deployment scenarios that focus on MU-MIMO. The advantage of OTFS in scaling capacity with the MIMO order makes it ideal for these kinds of deployments. In addition, the new radio air interface must support high spectral efficiency in high Doppler environments. OTFS is ideally suited for these requirements, providing: high spectral efficiency and reliability under diverse channel conditions.